\documentclass[a4paper, 11pt]{article}
\usepackage{geometry}
\usepackage{setspace}
\usepackage{amsmath}
\usepackage{amssymb}
\usepackage{graphicx}
\usepackage{booktabs}
\usepackage{dcolumn}
\usepackage{natbib}
\usepackage{subcaption}
\usepackage{xcolor} 

\usepackage{url}
\usepackage[colorlinks=true,citecolor=blue,linkcolor=blue,urlcolor=blue]{hyperref}
\usepackage{authblk}
\usepackage{footnote}
\usepackage{graphicx}
\usepackage{tikz}
\usepackage{bm}
\usepackage[bottom]{footmisc}
\usepackage{lineno}

\usepackage{algorithm}
\usepackage{algpseudocode}

\begin{document}

\title{\textbf{An ensemble prediction method for forecasting sap flux density and water-use in temperate trees}}
% Submitted to: Methods in Ecology and Evolution

\author[1]{Mengyi Gong\thanks{Corresponding E-mail: m.gong1@lancaster.ac.uk}}
\author[1,3]{Rebecca Killick}
\author[2]{Andrew Hirons}
\affil[1]{School of Mathematical Sciences, Lancaster University, Lancaster, U.K.} 
\affil[2]{Greenspace Department, University Centre Myerscough, Preston, U.K.}
\affil[3]{School of Mathematical and Statistical Sciences, Clemson University, South Carolina, USA}

\date{}
% \date{\today}

\maketitle

% The abstract should be numbered, not exceeding 350 words.
\abstract{
1. Efficient irrigation management is crucial to agriculture, forestry and horticulture, especially under climate change. Developments in novel sensors and Internet of Things technology provide an opportunity to carry out real-time monitoring of tree sap flux density, which, when coupled with advanced modelling techniques, enables online prediction of tree water-use suitable for irrigation planning. 

2. This manuscript proposes one such pipeline that integrates tree sap flow sensors, weather station sensors, and statistical models to predict tree daily water-use. In particular, an ensemble prediction approach based on additive models has been developed, using weather data as the main predictors of sap flux density. The method simultaneously considers the non-linear relationships and interactions between sap flux density and its environmental drivers, as well as the variability among individual trees over different growing seasons. 

3. Using field data collected on nine species of trees over the 2022, 2023 and 2024 growing seasons, this manuscript demonstrates the ability of the proposed ensemble prediction method in producing reliable daily water-use forecasts. The challenge of predicting tree water-use under climate stress, such as heatwaves, and the impact of tree sizes on prediction have been discussed.

4. Despite the complexity of the problem, the proposed method provides a general framework which can be used in a variety of settings, from commercial tree growers to conversation work. The model can be integrated into an online monitoring platform, assisting real-time decision making on irrigation management. 
}

\vspace{0.5cm}

% \begin{linenumbers}
\setstretch{1}

\section{Introduction}

Efficient irrigation protocols are essential to producing high-quality trees (and tree crops) whilst ensuring resilience to an increasingly warmer and drier climate. As tree water demand increases in response to the changing climate, irrigation in tree nurseries and orchards, needs to be scheduled more efficiently to maintain existing production levels and ensure future sustainability. Currently, irrigation scheduling decisions of tree nurseries are based almost entirely on manager experience without considering relevant data on soil, tree, or atmospheric water status, this often results in over-application of water. High irrigation volumes promote individual tree size but are inefficient, for it may limit overall nursery production as irrigation licences are capped at fixed volumes. More precise irrigation scheduling can maintain or even enhance tree quality by ensuring that substantial water deficits do not develop. 

Creating efficient irrigation protocols relies on new user-friendly tools that are appropriate for scaling up to production scale and are translatable across species and stock size. Developments in novel sensors and Internet of Things (IoT) technology provide an opportunity to carry out real-time monitoring of the behaviour of trees and the surrounding environment \citep{TreeTalker}. High-frequency time series data obtained from tree sap flux sensors, soil moisture sensors and weather sensors, when coupled with advanced statistical modelling techniques, have the potential to generate online prediction of tree water-use, allowing managers to make data-informed irrigation scheduling decisions in real-time. This manuscript develops such a novel approach to the prediction of tree water-use that can be scaled up to assist irrigation planning in agriculture, forestry or horticulture. % Should we expand this paragraph more?

Modern sap flow sensors sit at the centre of our approach. Time series data (often at sub-hourly frequencies) obtained from these sensors can provide a detailed picture of the temporal dynamics of tree sap flux densities. The modelling and prediction of sap flux density can be approached in several ways, from statistical methods, such as the ARIMAX($p, d, q$) (Integrated Auto-regressive Moving Average process with Exogenous variables) models \citep{TSAbook} which make use of the temporal dependence structure of the time series, and the semi-parametric regression models \citep{Regression} which make use of the complex relationships between the response and explanatory variables, to machine learning methods, such as random forest and neural networks \citep{MLprediction1, MLprediction2, MLprediction3}. Here we take the semi-parametric regression approach, namely the Generalized Additive Models (GAMs) \citep{GAM}, considering the features of the available data. Although the relationships between sap flux density and the key environmental variables, such as VPD and solar radiation, are well documented in literature \citep{SapfluxTropical, SapfluxPerhumid, SapfluxAsh, SapfluxDryForest}, the specific forms of these relationships are unclear, and they may vary from tree species to species. GAMs provide the flexibility of capturing different types of (potentially non-linear) relationships between variables. They also have the advantage over machine learning models in terms of interpretability. Visualising the relationships between the response and explanatory variables can provide useful insight into the scientific mechanics of the changes in sap flux density. To account for the natural variation among individual trees, we propose an ensemble prediction approach \citep{EnsembleNWP} where the GAMs fitted to individual trees are combined to produce a prediction for the species. We further propose a rolling prediction approach to make online prediction using data collected in real time, with the ensemble weights re-evaluated each time a new batch of data become available.

The prediction performance of the proposed ensemble prediction method is investigated in a case study using sap flux density data collected for nine species of trees (detail regarding the field data collection is given in Section \ref{sec:Data}) during the 2022, 2023 and 2024 growing seasons. The possibility of using the prediction model built on historical data to predict the sap flux density in a new growing season is also investigated, based on the data of five \textit{Tilia cordata} trees monitored over the 2023 and 2024 growing seasons. The case study illustrates good performance of the proposed method in predicting the sap flux density and water-use for the majority of the species, with the exception being the \textit{Pyrus calleryana} and the \textit{Carpinus betulus} during the 2022 summer heatwaves. This is discussed further following the case study, suggesting directions for future research.

The remainder of the manuscript consists of four sections. Section \ref{sec:Method} provides an introduction to the data collection procedure and the development of the ensemble approach for modelling and predicting tree sap flux density time series. Section \ref{sec:CaseStudy} presents a case study where the developed methods are applied to the sap flux density data collected for nine different species over three growing seasons, highlighting the advantages of the proposed method. Section \ref{sec:Discussion} discusses further the factors that can influence the prediction model, including heatwaves, tree size classes, and soil moisture levels, and the potential usage of the proposed methods in practice. Finally, section \ref{sec:Conclusion} concludes the manuscript.

\section{Materials and methods} \label{sec:Method}

\subsection{Data collection} \label{sec:Data}

A series of experimental field sites were established at Hillier Nurseries (Liss, Hampshire, UK) during the growing seasons 2022, 2023 and 2024. A weather station (air temperature, relative humidity, vapour pressure deficit (VPD), solar radiation, soil moisture and precipitation) was established at a central location, relative to the surrounding field sites. For each tree species of interest, five trees were selected for instrumentation: this included a soil moisture sensor (SoilMania, Arcen, NL) placed at a depth of 20-30 cm; a point dendrometer (ZN12-T-WP, Natkon, Wagen, CH) at between 1 - 1.5 m and a sap flow sensor (SFM1-X, ICT International, Armidale, AU) that used the heat ratio method to calculate sap flux density \citep{HeatPulseMethod1, HeatPulseMethod2}, installed 30 - 50 cm from ground level.  All sensors had IoT connectivity and were managed within a custom dashboard. 

A range of tree species, important to the urban landscape and the nursery trade, were selected for the study: 
Silver birch (\textit{Betula pendula}), hornbeam (\textit{Carpinus betulus}) and Callery pear (\textit{Pyrus calleryana}) were monitored during the 2022 growing season; rowan (\textit{Sorbus acuparia}), small-leaved lime (\textit{Tilia cordata}) and `New Horizon' elm (\textit{Ulmus} `New Horizon') were monitored during the 2023 growing season; and, red maple (\textit{Acer rubrum}), tulip tree (\textit{Liriodendron tulipifera}), whitebeam (\textit{Sorbus aria}) and \textit{T. cordata} during the 2024 growing season. Some species also had different size classes based on their circumference (in cm) at a height of 1 metre. Across the experiment, these size classes ranged from `standard' (10-12) to `semi-mature' (25-30) \citep{BS}. In total, 12 groups of trees from nine different species were investigated (see Table \ref{tab:EnsembleExperiment}). 

From sap flux density, the water-use of a tree can be estimated by multiplying by the total sapwood area. Denote the sap flux density at time $t$ as $Y_{t}$. First, calculate the sapwood area based on the circumference (denoted as $c$) and bark depth (denoted as $d$) as $A = \pi (c/2\pi - d)^{2}$. Then the water-use between time $t_{0}$ and $t_{n}$ can be obtained as $\sum_{t=t_{0}}^{t_{n}}AY_{t}\Delta_{t}$, where $\Delta_{t}$ is the gap between two adjacent time points measured in hours. For larger trees, multiple sensors are sometimes installed at different depths of the tree trunk to obtain more accurate measurements. If a set of two sensors were installed at two different depths of the tree trunk with the inner sensor measuring an area of radius $r_{1}$ from the center, with measurement $Y_{t1}$, and the outer sensor measuring the area between radius $r_{1}$ and $r_{2} = c/2\pi - d$, with measurement $Y_{t2}$. The sapwood area covered by the inner sensor would be $A_{1} = \pi r_{1}^2$ and the sapwood area covered by the outer sensor would be $A_{2} = \pi (r_{2}^{2} - r_{1}^{2})$. Hence the water-use can be estimated as $\sum_{t=t_{0}}^{t_{n}}A_{1}Y_{t1}\Delta_{t} + \sum_{t=t_{0}}^{t_{n}}A_{2}Y_{t2}\Delta_{t}$. Alternatively, one can take the average of the sap flux density time series measured at different depths and obtain the water-use using the averaged sap flux density as $\sum_{t=t_{0}}^{t_{n}} A [(Y_{t1} + Y_{2t}) / 2] \Delta_{t}$.

The above estimation method assumes that the measured sap flux density is uniform to the corresponding sapwood area, which is often deemed as over-simplified. However, modelling the radial sap flux density is itself a challenging task \citep{SapfluxSpatial, RadialSapFlux}, and the proposed solutions each have different assumptions, advantages, and disadvantages. Hence, it is not considered in this study. For the field set up at Hillier Nurseries, a set of two thermocouples, positioned 15 mm apart on the sensor needle, were installed at two different depths within the tree stem, the outer thermocouple was positioned 5 mm into the sapwood (xylem) and the inner thermocouple 20 mm into the sapwood. The two thermal couples recorded the outer and inner sap flux densities respectively. In what follows, the sap flux density of the outer ring is used to estimate the water-use of smaller trees (with radius smaller than 3.5 cm) and the averaged sap flux density from the inner and outer ring is used to estimate the water-use of larger trees (with radius greater than 3.5 cm).

\begin{figure}[!htb]
\begin{center}
\begin{minipage}{3.2in}
\includegraphics[width=3.1in]{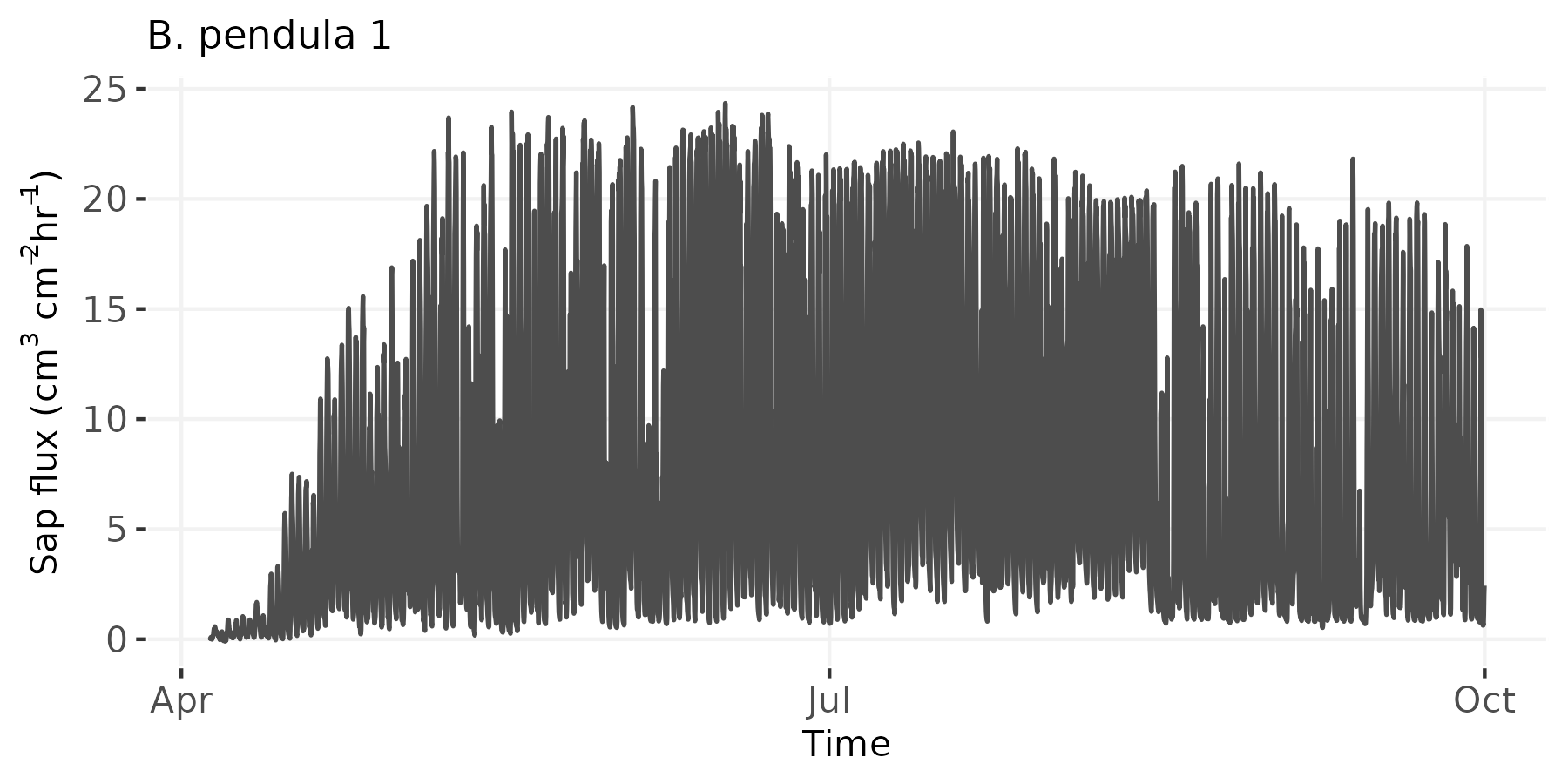}
\includegraphics[width=3.1in]{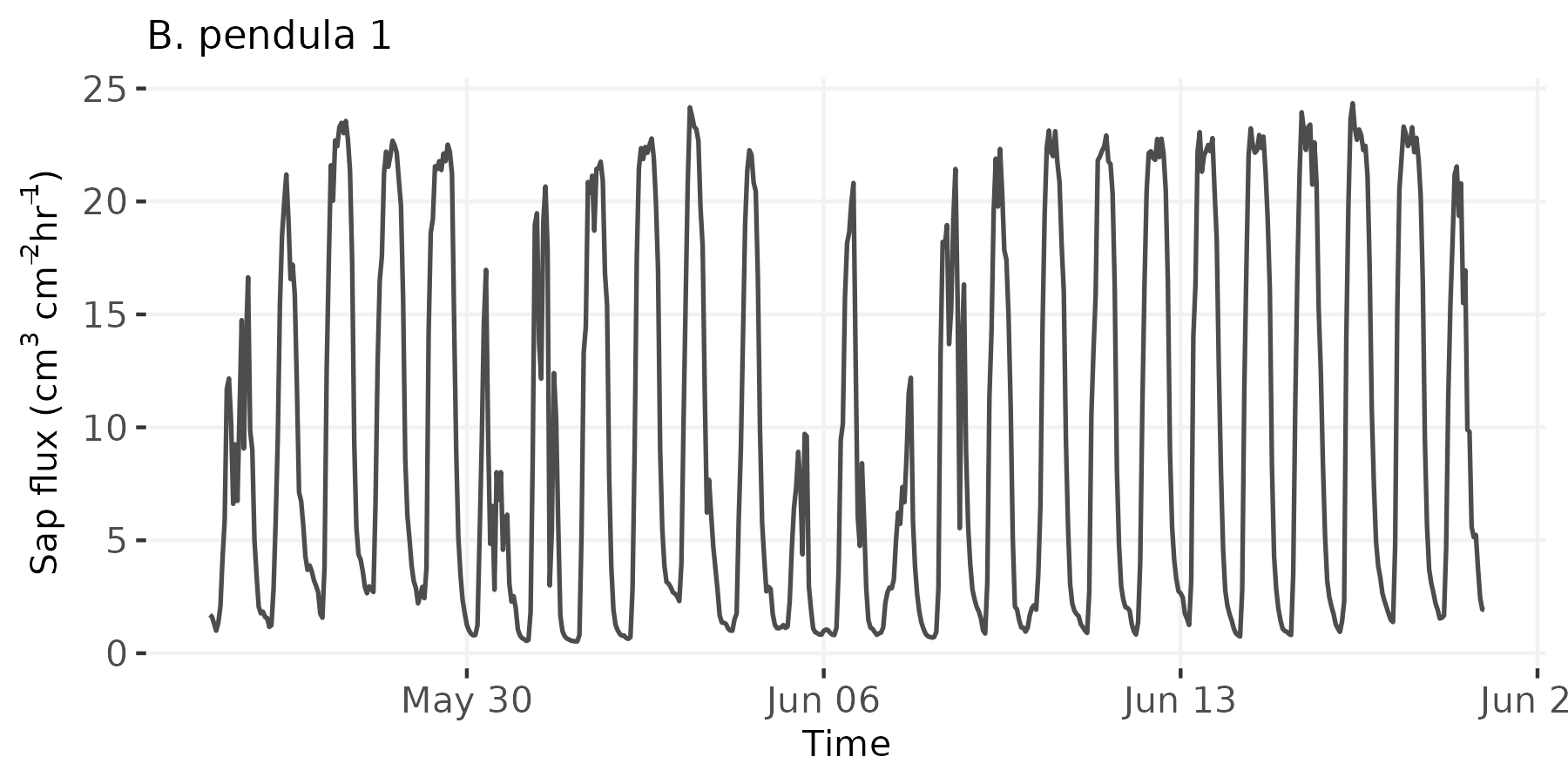}
\end{minipage} %
\begin{minipage}{3.2in}
\includegraphics[width=3.1in]{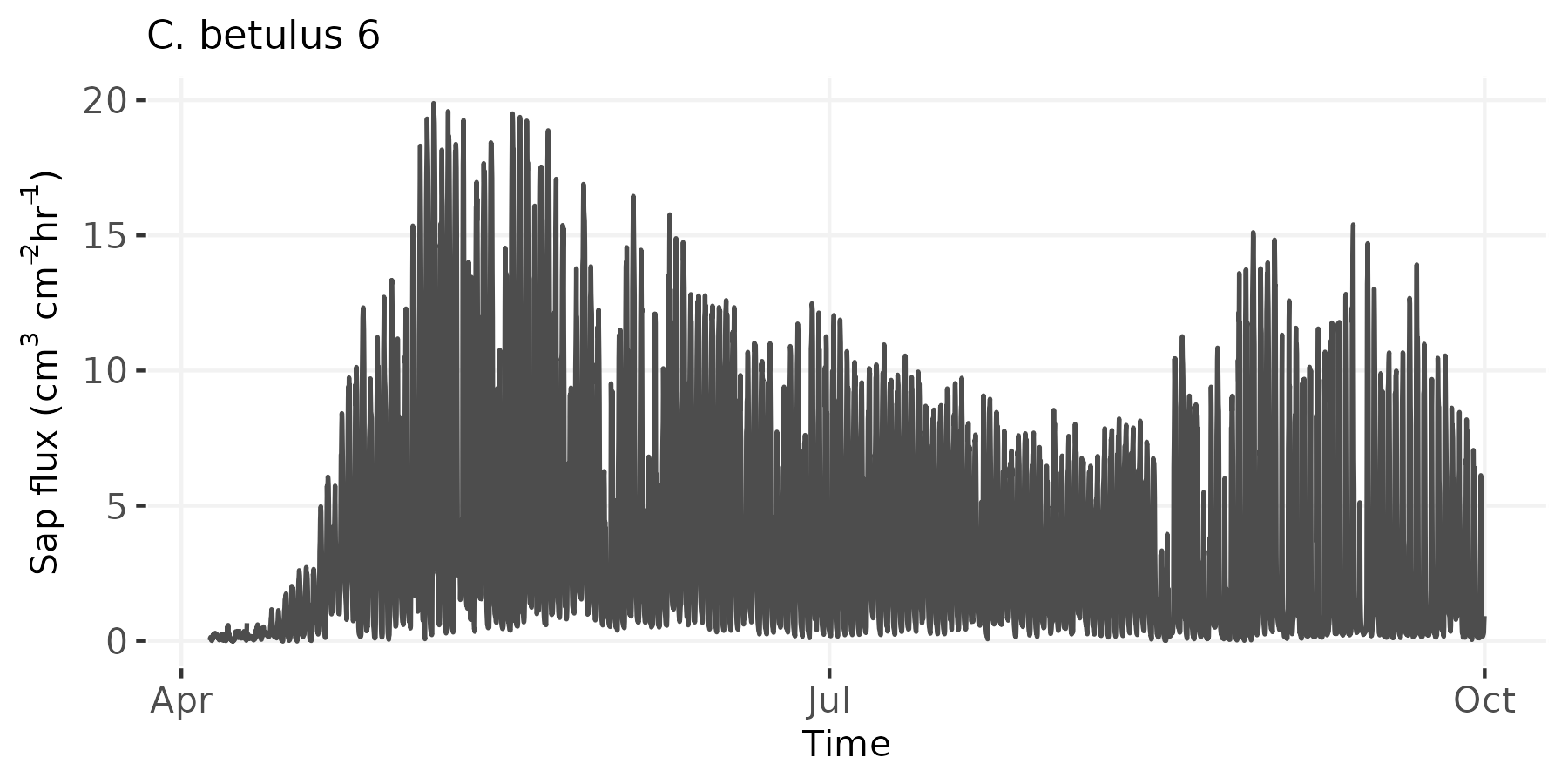}
\includegraphics[width=3.1in]{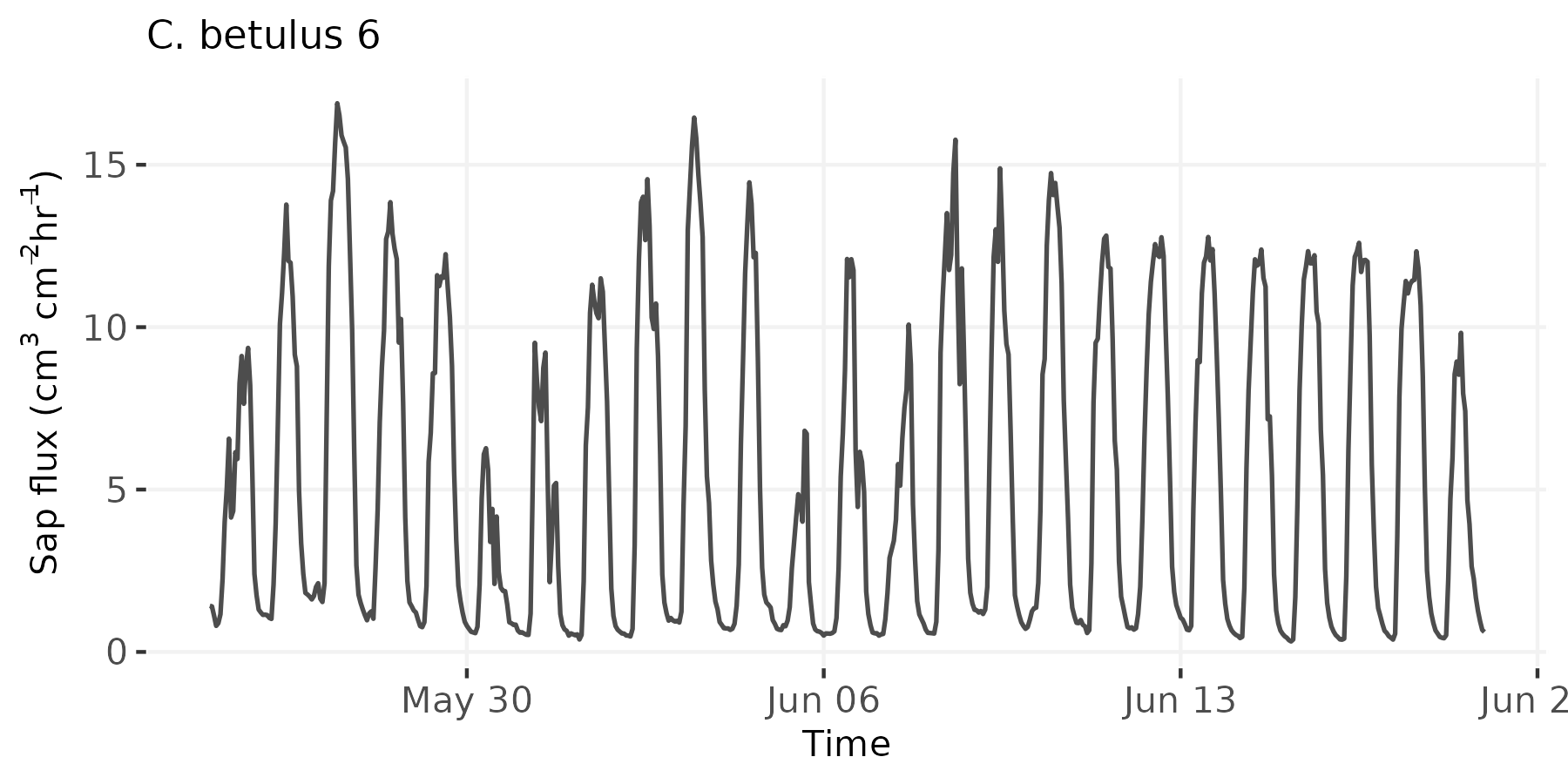}
\end{minipage}
\captionsetup{labelfont=bf, font=small}
\caption{Time series plot of sap flux density for \textit{B. pendula} 1 and \textit{C. betulus} 6. The top panels show the time series of the entire growing season from April to September 2022. The bottom panels show a shorter period between 25 May to 18 June 2022.}
\label{fig:ts_Betula}
\end{center}
\end{figure}

Figure \ref{fig:ts_Betula} shows examples of the hourly sap flux density time series of a \textit{Betula pendula} tree (labelled \textit{B. pendula} 1) and a \textit{Carpinus betulus} tree (labelled \textit{C. betulus} 6) from the 2022 growing season. There are clear daily cycles in the time series, where the sap flux peaks at around mid-day to early afternoon and drops to 0 at around mid-night to early morning. Initially, the scales of the daily cycle followed an increasing trend in early spring when the new leaves are expanding, then they started to fluctuate after the leaves were fully expanded. The temporal patterns of the scales of daily cycles varied from tree to tree, with the differences between species being more distinctive. For \textit{B. pendula} 1, the scales of the daily cycles stayed at around the same level in July, decreased in August and then increased again in September. For \textit{C. betulus} 6, there was a significant drop of daily maximum sap flux density in July and Auguts, before the scale of daily cycles rose again. These patterns may be explained by the persistent hot and dry weather in July and August, followed by the cool and rainy period in September. They seem to reflect the differences in the response of individual trees to the changing weather conditions.  

\subsection{Statistical modelling} \label{sec:Stats}

\subsubsection{Exploring the relationships between variables} \label{sec:Relationship}

As an important part of the exploratory analysis, time series data of the sap flux density and the environmental variables were plotted in various ways to investigate the relationships between variables (e.g., linear versus non-linear) and the auto-correlations / cross-correlations between observations at different time lags. The following discussion focuses on the scatter plots between VPD, solar radiation, air temperature, relative humidity and sap flux density. These plots provide unique insight into the interactions between sap flux density and its environmental drivers, which is crucial to the construction of the prediction model. 

\begin{figure}[!htb]
\begin{center}
\includegraphics[width=6.5in]{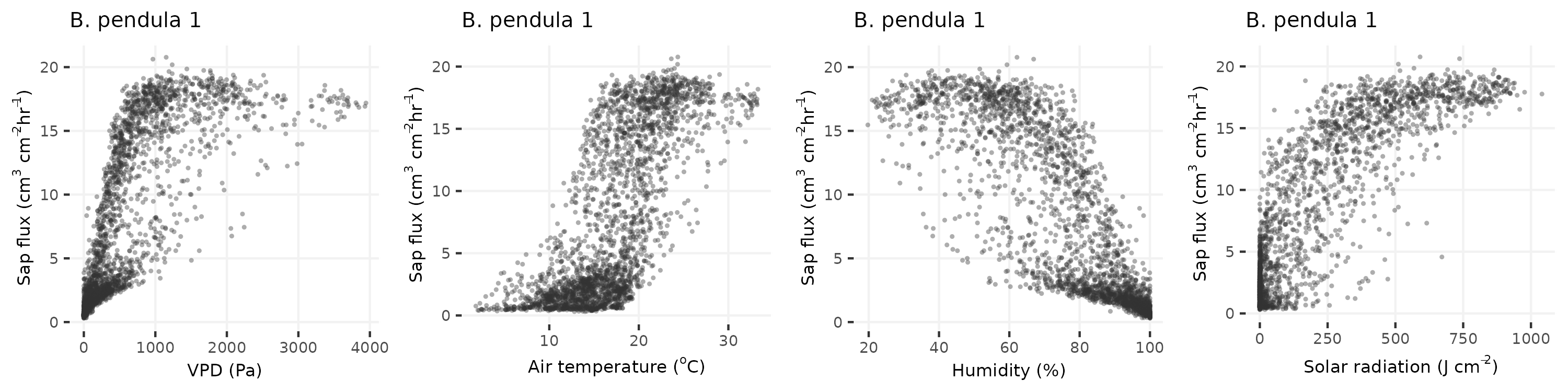}
\includegraphics[width=6.5in]{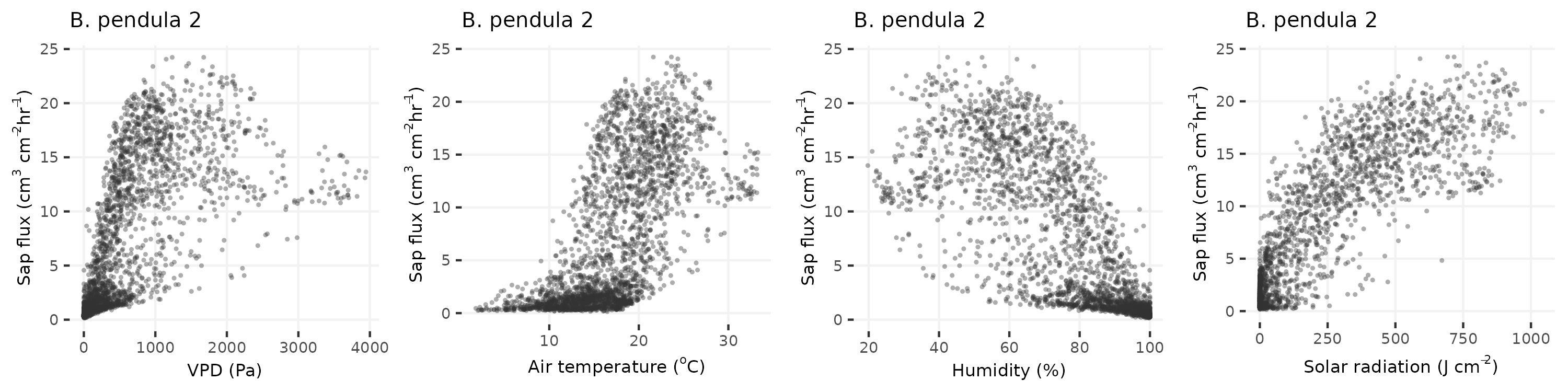}
\captionsetup{labelfont=bf, font=small}
\caption{Scatter plots between sap flux density and VPD, air temperature, humidity and solar radiation for two \textit{B. pendula} trees.}
\label{fig:scatter_nonlinear_Betula}
\end{center}
\end{figure}

\begin{figure}[!htb]
\begin{center}
\includegraphics[width=5.5in]{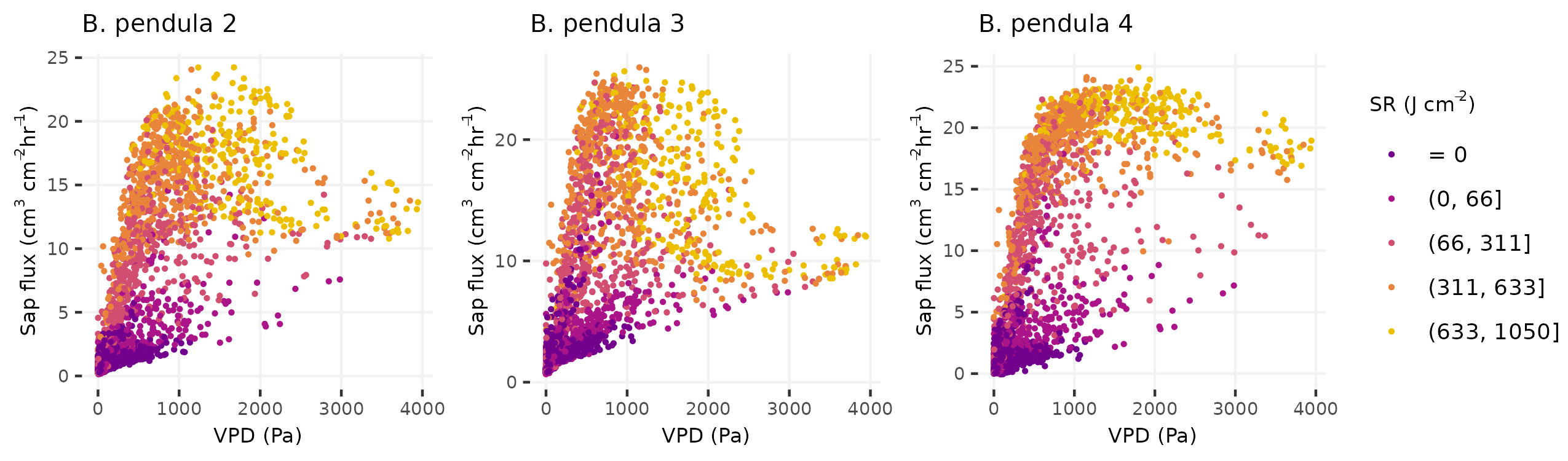}
\captionsetup{labelfont=bf, font=small}
\caption{The scatter plot between sap flux density and VPD for three \textit{B. pendula} trees, where the points are coloured by five levels of solar radiation (SR), with the purple end of the palette representing low solar radiation and the yellow end representing high solar radiation.}
\label{fig:scatter_interact_Betula}
\end{center}
\end{figure}

Figure \ref{fig:scatter_nonlinear_Betula} shows an example of the scatter plots between the hourly sap flux density of two \textit{Betula pendula} trees and the environmental variables from the 2022 growing season. The scatter plots indicate clear non-linear relationships between sap flux density and the environmental variables, in particular with VPD and solar radiation. This strengthens the non-linear relationships noted in \cite{EnvControlSapflux}, \cite{SapfluxDrivers} and \cite{ SapfluxCherry}. To further explore the interactions between variables, different colour schemes were used to produce the scatter plots to help visualise the inter-relationships. Figure \ref{fig:scatter_interact_Betula} shows the scatter plots between sap flux density and VPD, coloured by five different levels of solar radiation for three \textit{B. pendula} trees. The plots show clear divergence in the relationships from the low to high solar radiation periods. A similar phenomenon has been discussed in \cite{SapfluxTropical} where the VPD and irradiance co-vary with sap flux density, and in \cite{SapfluxRadiation} where the relationship between sap flux density and other environmental variables can change under different radiation levels. These results suggest the use of an interaction term between VPD and solar radiation in models. 

Note that as we used hourly time series in this study, it is possible to investigate the sap flux density lagged dependence with environmental variables (e.g., solar radiation) via hysteresis graphs as in \cite{LagDependence1} and \cite{LagDependence2} to better characterise the relationship between variables. Alternatively, cross-correlation plots, which shows the correlation between two variables with certain time lag $\tau$, i.e., $Corr(X_{t}, Y_{t+\tau})$, can also help identify the lagged dependence between variables. Such analysis could improve the understanding of the mechanics of changes in sap flux density. In this case, however, identifying the lags between variables brings little improvement to the prediction of sap flux density. Therefore, the lagged dependence is not considered in the following content. Further explanation on the use of lagged covariates and its performance is given in the supplementary document.

\subsubsection{An additive model for sap flux density of individual trees} \label{sec:Additive}

Motivated by the features learned from the exploratory analysis, we propose to use a flexible regression approach, the generalised additive model (GAM), to capture the relationships between the response and explanatory variables. GAMs have the advantage of being flexible and can account for arbitrary non-linear relationship whose parametric form is difficult to obtain, which happens to be the case with the relationships between sap flux density and the environmental variables.

The general structure of the additive model for the sap flux density of an individual tree is assumed to be
\begin{equation}
Y_{t} = \alpha_{0} + \alpha_{1} Y_{t-1} + \alpha_{2} T_{t} + \alpha_{3} H_{t} + \bm{s}_{1}(R_{t})^{\top} \bm{\beta}_{1} + ( \bm{s}_{2}(V_{t}) \cdot R_{t} )^{\top} \bm{\beta}_{2} + \bm{s}_{3}(V_{t} , X_{t})^{\top} \bm{\beta}_{3} + \epsilon_{t} \; ,
\label{eqn:GAM_individual}
\end{equation}
where $Y_{t}$ represents the sap flux density at time $t$, $Y_{t-1}$ is the lag-1 sap flux density used to account for the auto-correlations, $T_{t}$, $H_{t}$, $R_{t}$ and $V_{t}$ are air temperature, relative humidity, solar radiation and VPD at time $t$, and $X_{t}$ is a flexible variable that is to be determined based on species and growing seasons. The model residual $\epsilon_{t}$ is assumed to be independently and identically distributed from a Normal distribution. Explicitly speaking, VPD and solar radiation are often considered as the key factors driving sap flux density in literature \citep{EnvControlSapflux, SapfluxDrivers}, and hence they are used as the main predictors in the model. Based on the observed non-linear correlations and interactions in the scatter plots, smooth functions of $R_{t}$ and $V_{t}$ are used to capture the non-trivial inter-relationships between the variables. These smooth functions are constructed using a set of basis functions, $\bm{s}(\cdot) = ( s_{1}(\cdot), \ldots, s_{p}(\cdot) )^{\top}$, and the corresponding basis coefficients, $\bm{\beta}= (\beta_{1}, \ldots, \beta_{p})^{\top}$, which are estimated together with other model coefficients. Specifically, the additive term $\bm{s}_{1}(R_{t})^{\top} \bm{\beta}_{1}$ describes the non-linear relationship between sap flux density and solar radiation. The additive term $( \bm{s}_{2}(V_{t}) \cdot R_{t} )^{\top} \bm{\beta}_{2}$ is used to describe the non-linear relationship between sap flux density and VPD that varies with different levels of solar radiation (recall the interactions displayed in Figure \ref{fig:scatter_interact_Betula}). Finally, the interaction term $\bm{s}_{3}(V_{t}, X_{t})$ is included as a flexible component to capture the variation in the scales of daily cycles of the sap flux density time series. As described in Section \ref{sec:Data}, even after the leaves are fully expanded, there are still temporal trends in the scales of daily cycles over the growing season. These temporal trends are non-trivial in that none of the available environmental variables can fully explain the fluctuations, although some may share similar patterns to the changing scales of a particular species in a particular growing season. Hence, the flexible term $\bm{s}_{3}(V_{t}, X_{t})$ is used, where the choice of $X_{t}$ and the type of the interaction (i.e., whether it is a simple interaction term, a varying coefficient term or a bi-variate smooth term) are left to decide upon the features of the time series under investigation. 

\begin{figure}[!htb]
\begin{center}
\includegraphics[width=4.8in]{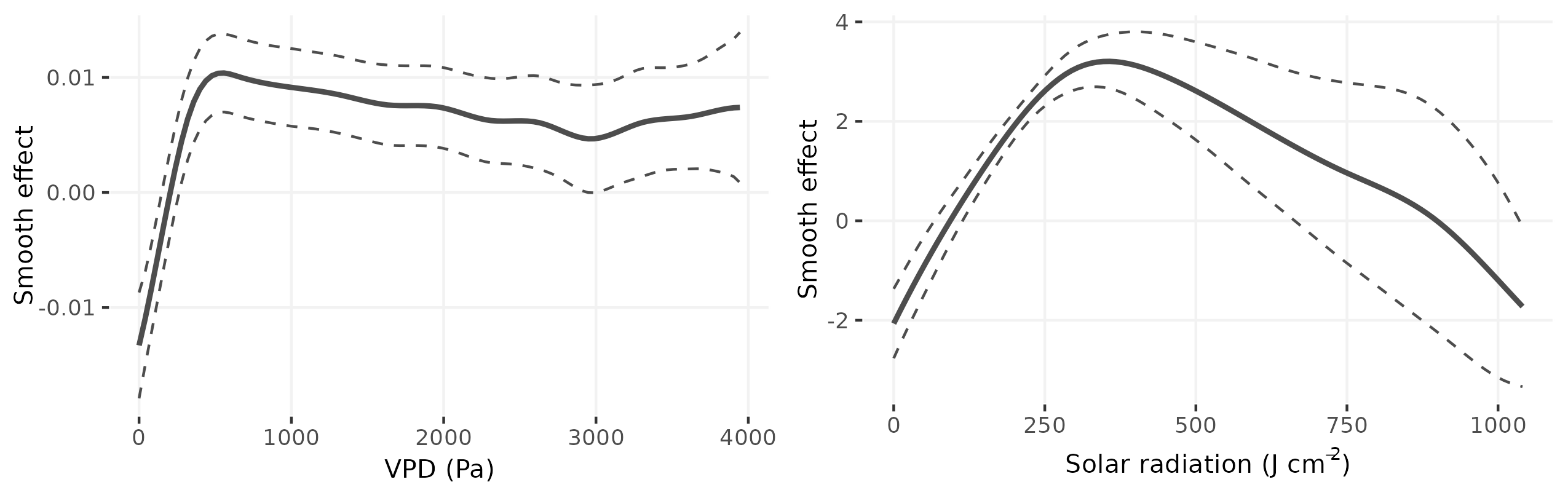}
\captionsetup{labelfont=bf, font=small}
\caption{Estimated smooth non-linear effect of VPD (left) and solar radiation (right) from the additive model (\ref{eqn:GAM_individual}) fitted to the sap flux time series data of \textit{B. pendula} 1.}
\label{fig:Betula_gam}
\end{center}
\end{figure}

Model (\ref{eqn:GAM_individual}) can be fitted using the \texttt{mgcv} package in \texttt{R} \citep{mgcv}. As an illustration, we present the result from fitting the additive model (\ref{eqn:GAM_individual}) to the sap flux density time series of \textit{B. pendula} 1 during the 2022 growing season. Following the exploratory analysis, the variable $X_{t}$ in the flexible term $\bm{s}_{3}(V_{t} , X_{t})$ is taken to be daily maximum temperature. The fitted model has a high deviance explained (97.8\%), and the residuals are mostly small with a median of 0.3701. The auto-correlations of different lags in the residual time series are generally insignificant, apart from the moderate auto-correlation at lag 24. The estimated smooth non-linear effect of VPD and that of solar radiation are visualised in Figure \ref{fig:Betula_gam}, where both variables show an increasing trend at lower values which then turns into a flat or even decreasing trend at higher values. This appears to be consistent with the findings (with respect to a few different species) in the literature where the solar radiation drives the increase of sap flux density initially, before its impact plateaus or even decreases at higher levels \citep{SapfluxCherry, NeedleSensor}.  The sap flux density increases with VPD initially, before the increment becomes negligible or negative, potentially due to stomatal closure \citep{SapfluxBirch, EnvControlSapflux, NeedleSensor}. Since the 2022 growing season was a hot and dry summer with two significant heatwaves, the decreasing patterns at high solar radiation and VPD levels are to be expected.

Model (\ref{eqn:GAM_individual}) can be used to make $k$-step ahead predictions of sap flux density in the future. To do this, weather forecast data for the prediction period are required. Prediction is calculated using a rolling one-step ahead prediction, starting from the current time. This is due to the presence of the lagged predictor, i.e., the term $Y_{t-1}$ in the model (\ref{eqn:GAM_individual}), which prohibits batch calculation. Examples of the predicted sap flux density time series are given in the supplementary material.

\subsubsection{Ensemble prediction of sap flux density of a group of trees} \label{sec:Ensemble}

Despite the similarity in the non-linear patterns in general, the scatter plots and the time series plots show variations across individual trees, even if they belong to the same species or the same size class. Therefore, it is inappropriate to use any individual tree to represent an entire group of trees. However, by pooling the information from all monitored trees from the same group, an estimation more representative of the population mean may be achieved, provided that the trees being monitored do not behave drastically different from the rest of the population. Motivated by this, an ensemble prediction approach has been proposed to forecast the sap flux density and hence the water-use of a group of trees. The main idea is to combine the predictions from the additive models built for individual trees to obtain the prediction of the sap flux density of a `standard tree'. Then estimate the water-use of a `standard tree' and scale up by the total number of trees to get the group water-use.

Denote the sap flux density model $m$ built on the data of tree $m$ as $f_{m}$, and the prediction of sap flux density from model $m$ as $\widehat{Y}_{t}^{m}$. Each $f_{m}$ is considered as an ensemble element or ensemble member, and each ensemble element is assigned an ensemble weight $w_{m}$, subject to $\sum_{m=1}^{M} w_{m} = 1$. The ensemble prediction of the sap flux density is then obtained as, 
\begin{equation}
\widehat{Y}_{t} = \sum_{m=1}^{M} w_{m} \widehat{Y}_{t}^{m} \; .
\label{eqn:EnsemblePred}
\end{equation}
The ensemble weights are usually obtained based on training data, which in this case would be the observed sap flux density data from a period in the past. Some commonly used ensemble weights include (a) equal weights, (b) reciprocal of mean squared errors (MSE) weights, and (c) penalised regression weights, which takes the predictors $\widehat{Y}_{t}^{m}$ as regressors and fit the linear regression model with lasso, ridge or elastic net penalty \citep{StatsLearning}. More sophisticated weighting schemes are available, such as in \cite{AdditiveStack}. Although in this case, the equal weights and the reciprocal of MSE weights appear to perform better. Some discussion on the performance of different weights can be found in \cite{ForecastAvg}.

The ensemble can be further expanded by including different models for each monitored tree. This was motivated by the flexible choices of the variable $X_{t}$ when modelling the sap flux density data from different species over different growing seasons. From the available data, we identified three candidates for $X_{t}$, which are daily maximum air temperature, daily minimum relative humidity and daily mean soil moisture. From initial experiment, it appears that, for a particular tree, one variable may be superior to the others in predicting sap flux density. However, such a superiority is not always consistent within a species or a group, and it may change from one period to another. It raises the question of whether it is sensible to spend time and effort determining the most appropriate $X_{t}$ for each individual tree. A simple alternative is to consider all candidate models as ensemble members, and produce the ensemble prediction via appropriate weighting. If the ensemble weights are updated when the prediction is made, then the model with the most appropriate $X_{t}$ for the prediction period is likely to receive a higher weight, reflecting the preference of the choice over the flexible variable. In general, as more data are collected and more ensemble members become available, it is more likely that some of them will be suitable for the prediction problem at hand. 

A typical prediction task for a tree nursery or an orchard is to predict the water-use of a group of trees in the next few days or weeks, based on the weather forecast data. For this task, a rolling ensemble prediction approach has been proposed to achieve better prediction performance and computational efficiency. In a nutshell, the rolling ensemble prediction iteratively updates the additive models (i.e., ensemble members) based on data up to the current time, re-trains the ensemble weights based on the data from the most recent period, makes prediction of future sap flux density of a `standard tree' using the updated ensemble members and weights, and finally forecasts the water-use using the predicted sap flux density. Algorithm \ref{alg:Ensemble} shows the key steps of the rolling ensemble prediction procedure. It assumes that $n$ trees are monitored, $q$ types of models are used and the water-use of $N$ trees is to be estimated. The prediction is carried out on the next $d$ time points. Although there is one set of weather forecast data, the use of lag-1 sap flux density in the prediction model naturally creates $n$ different prediction data sets, each with a different initial condition $Y_{(t-1)i}$, $i = 1, \cdots, n$, from $n$ observed trees. For each prediction data set $i$, an ensemble mean $\widehat{Y}_{ti}$ can be obtained, and the final prediction of the `standard tree' can be obtained by averaging $n$ ensemble mean, as  
\begin{equation}
    \widehat{Y}_{t}^{\ast} = \frac{1}{n} \sum_{i=1}^{n} \widehat{Y}_{ti} = \frac{1}{n} \sum_{i=1}^{n} \left( \sum_{m} w_{m} \widehat{Y}_{ti}^{m} \right) \; .
    \label{eqn:StandardTree}
\end{equation} 
This approach can help taking into account the uncertainty in initial conditions. A similar approach can be taken when there are several competing weather forecast data from different sources, in which case, the data set index $i$ becomes the index of the forecast sources. This can help mitigate the uncertainty in the forecast data.

\begin{algorithm}[htb]
\caption{Rolling ensemble prediction}\label{alg:Ensemble}
\begin{algorithmic}
\Require Sap flux data from $n$ trees $Y_{ti}$, $i = 1, \cdots, n$; weather forecast data $\bm{X}_{t}$; $q$ different types of models used in the ensemble prediction; the initial period of $d_{0}$ time points; the rolling prediction period of $d$ time points   \\
\For{$t = d_{0}, d_{0} + d, d_{0} + 2d, \cdots$, }  
  \State (1). Fit all $q$ candidate models to data up to time $t$. This gives us $qn$ ensemble members. Denote the models as $f_{m}$, $m=1, \cdots, qn$.
  \State (2). Apply the models to each training data set $i$, $i=1, \cdots, n$ . Denote the prediction from model $f_{m}$ as $\widehat{Y}_{(t+1)i}^{m}, \cdots, \widehat{Y}_{(t+d)i}^{m}$. 
  \State (3). Obtain the prediction error as $\epsilon_{(t+s)i}^{m} = \widehat{Y}_{(t+s)i}^{m} - Y_{(t+s)i}$, for $s=1, \cdots, d$. Use the prediction errors or related quantities to obtain the ensemble weights, denoted as $w_{m}$. 
  \State (4). Use the models and weights to make ensemble prediction for the next $d$ time points based on prediction data set $i$, $i = 1, \cdots, n$. Denote the prediction as 
    \begin{equation*}
        \widehat{Y}_{(t+s)i} = \sum_{m} w_{m} \widehat{Y}_{(t+s)i}^{m} \; , \; \; s=1, \cdots, d \; .
    \end{equation*} 
  \State (5). Generate the prediction of a `standard tree' by averaging the ensemble means,
    \begin{equation*}
    \widehat{Y}_{(t+s)}^{\ast} = \frac{1}{n} \sum_{i=1}^{n} \widehat{Y}_{(t+s)i} = \frac{1}{n} \sum_{i=1}^{n} \left( \sum_{m} w_{m} \widehat{Y}_{(t+s)i}^{m} \right) \; , \; \; s=1, \cdots, d \; .
    \end{equation*} 
  \State (6). Estimate water-use by scaling up the total sapwood area of $N$ ($N \geq n$) trees, 
    \begin{equation*}
    \widehat{Z}_{t+s} = A^{\ast} \widehat{Y}_{t+s}^{\ast} = \left( \sum_{j=1}^{N} A_{j} \right) \widehat{Y}_{t+s}^{\ast} \; , \; \; s=1, \cdots, d \; .
    \end{equation*} 
\EndFor

\hspace{-7ex}\Return A set of predicted water-use $\widehat{Z}_{t+1}, \cdots, \widehat{Z}_{t+d}$. % $\{\widehat{Z}\}_{t+1, \cdots, t+d}$
\end{algorithmic}
\end{algorithm}

The uncertainty of ensemble prediction is typically evaluated through investigating the relationship between the ensemble spread and the prediction error \citep{EnsembleSpread}. There are different ways of defining the ensemble spread. One of them is the averaged squared deviation between the ensemble prediction and the prediction from ensemble members,  
\begin{equation}
\mathcal{S}_{t} = \sqrt{ \sum_{m} w_{m} \left( \widehat{Y}_{t}^{m} - \widehat{Y}_{t} \right)^{2} } \; ,
\label{eqn:Spread}
\end{equation}
where $\mathcal{S}_{t}^{2}$ is sometimes referred to as the variance of the ensemble prediction. The average of the absolute deviation can be another choice for ensemble spread. The relationship between the ensemble spread $\mathcal{S}_{t}$ and the ensemble error $\mathcal{E}_{t}$ is usually established using some training data. This relationship, whether it is linear or non-linear, is then used to estimate the ensemble prediction errors from the ensemble spread of future predictions. For example, if a linear relationship is established, then one can scale the ensemble spread by a scaling factor $\gamma$ to obtain the ensemble error \citep{EnsembleError} as $\mathcal{E}_{t} = \gamma \, \mathcal{S}_{t}$. The scaling factor can be estimated using training data as,
\begin{equation}
    \gamma = \sqrt{ \frac{\sum_{t} \sum_{m} w_{m} \left( \widehat{Y}_{t}^{m} - Y_{t} \right)^{2} }{ \sum_{t}\mathcal{S}_{t}^{2} } } \; .
    \label{eqn:SpreadScale}
\end{equation} 

The last step in obtaining the uncertainty measure is error propagation. This is required due to the scaling up step used to estimate the water-use of a group of trees. As the scaling is a linear operation, simple linear error propagation can be applied to get the final uncertainty measure. Note that this is based on the assumptions that the monitored trees are fair representations of the trees in the group. This may not be the case when it comes to a large number of trees from a vast field with changing landscape. Some adjustment may be needed to better reflect the uncertainty.

\section{Results from the case study} \label{sec:CaseStudy}

\subsection{Ensemble prediction of a growing season}

In the first illustration, the historical data collected from Hillier Nurseries during the 2022, 2023 and 2024 growing seasons are used to mimic a situation where the sap flux density data are obtained in real time, and the rolling ensemble prediction is carried out by treating the weather data and soil moisture data as `pseudo' forecast data. 

The additive model (\ref{eqn:GAM_individual}) was fitted to the trees that have sap flux density time series without major missing gaps. Considering overall data availability, the time period used for the ensemble prediction experiment are 1 July 2022 to 15 September 2022 for \textit{B. pendula}, \textit{C. betulus} and \textit{P. calleryana}, 15 June 2023 to 30 August 2023 for \textit{T. cordata}, 1 July 2023 to 30 August 2023 for \textit{S. aucuparia} and \textit{U.} `New Horizon', and 1 June 2024 to 15 September 2024 for \textit{A. rubrum}, \textit{L. tulipifera}, and \textit{S. aria}. The prediction started from the third week of the selected time period (i.e., there are at least two weeks of training data) and a rolling window of seven days was used throughout the experiment. Iteratively, at the end of the current 7-day window, the additive models were updated with data up to current time, the associated ensemble weights (which were the reciprocal of MSE weights) were recalculated and the prediction of sap flux density of the next seven days was made using the ensemble mean. The daily water-use predictions were then obtained by multiplying the predicted sap flux density time series with the corresponding sap wood areas of the trees, and aggregated by day. 

Some measures on prediction performance are presented in Table \ref{tab:EnsembleExperiment}. As the scales of sap flux density and the water-use differ significantly from group to group, it is inappropriate to compare the prediction errors directly. Therefore, the percentage prediction errors, which are the absolute prediction error divided by the observed sap flux density or water-use, were calculated for comparison. In particular, we presented in Table \ref{tab:EnsembleExperiment} the MSE of the sap flux density prediction, medium length of the half confidence interval of the sap flux density prediction and the 0.1, 0.5 and 0.9 quantiles of the percentage prediction errors of water-use over the experimental period. Figures showing the ensemble prediction of the sap flux density and daily water-use of a standard tree for the `16-18' \textit{P. calleryana} in 2022 and the `18-20' \textit{T. cordata} in 2023 are given in the supplementary document.

\begin{table}[htb]
\captionsetup{labelfont=bf, font=small}
\caption{Summary of the MSE, medium length of the half confidence interval (Half CI) of the prediction of hourly sap flux density and the 0.1, 0.5 and 0.9 quantiles of the percentage errors (Err\%) of the prediction of daily water-use from the ensemble prediction experiment carried out on nine species.} \label{tab:EnsembleExperiment}
\centering
\small
\begin{tabular}{l|cc|ccc}
  \hline
  \textbf{\textit{Species} `size-class' (Year)} &  \multicolumn{2}{c}{\textbf{Hourly sap flux}} & \multicolumn{3}{c}{\textbf{Daily water-use}} \\
  & MSE &  Half CI (0.5) &  Err\% (0.1) & Err\% (0.5) & Err\% (0.9) \\ 
  \hline
  \textit{B. pendula} `25-30' (2022) & 2.56 & 1.56 & 1.53 & 9.46 & 17.86 \\ 
  \textit{C. betulus} `20-25' (2022) & 1.49 & 0.94 & 2.64 & 8.32 & 22.88 \\ 
  \textit{C. betulus} `16-18' (2022) & 5.33 & 0.76 & 5.39 & 23.45 & 58.58 \\ 
  \textit{P. calleryana} `16-18' (2022) & 1.77 & 0.83 & 12.42 & 31.50 & 56.07 \\ 
  \textit{S. aucuparia} `12-14' (2023) & 0.49 & 0.55 & 1.04 & 5.76 & 17.14 \\  
  \textit{T. cordata} `18-20' (2023) & 1.86 & 0.79 & 1.79 & 8.74 & 19.18 \\ % Youngs 
  \textit{T. cordata} `12-14' (2023) & 3.40 & 1.66 & 1.06 & 6.10 & 26.48 \\ % Adhurst
  \textit{U.} `New Horizon' `20-25' (2023) & 0.35 & 0.65 & 1.77 & 8.81 & 20.48 \\ % Westmark North
  \textit{U.} `New Horizon' `12-14' (2023) & 0.36 & 0.59 & 1.21 & 7.03 & 20.31 \\ % Andlers Ash
  \textit{A. rubrum} `12-14' (2024) & 0.81 & 0.88 & 2.82 & 12.12 & 25.34 \\
  \textit{L. tulipifera} `14-16' (2024) & 1.80 & 1.14 & 2.09 & 11.69 & 36.65 \\
  \textit{S. aria} '10-12' (2024) & 0.87 & 0.68 & 4.21 & 14.68 & 28.28 \\
   \hline
\end{tabular}
\end{table}

Overall, the rolling ensemble prediction performed well in forecasting sap flux density and daily water-use. Among the 12 groups, most of the groups have a medium percentage prediction error between 10\% and 20\% for hourly sap flux density and below 10\% for daily water-use, expect for the \textit{P. calleryana} group and the \textit{C. betulus} `16-18' group, where the percentage prediction errors are distinctively higher. In fact, the \textit{P. calleryana} group has the highest percentage prediction errors in all the quantiles presented in the table for hourly sap flux density and daily water-use. A closer inspection of the predicted sap flux time series of the \textit{P. calleryana} group showed large discrepancy between the predictions and the observations during the two heatwaves and the post heatwave period in 2022. In general, the ensemble prediction approach appears to work better in the 2023 and 2024 growing seasons, which were typical English summers, than the 2022 growing season, which had record high temperatures. 

One explanation for the discrepancies between the predicted and observed sap flux during and after heatwave events is that the decline in soil water availability during a period of high VPD resulted in water potentials that induced embolism in the xylem vessels \citep{EmbolismResistance, XylemEmbolism}. The concomitant reduction of hydraulic conductivity depressed the amplitude in sap flux density, even after soil moisture was recharged following rain (see the supplementary document for an example of \textit{P. calleryana}). Over time, ensemble models with additional seasons' data, are expected to reduced the uncertainty associated with stress events.    

\subsection{Using historical models to predict future water-use}

Another typical prediction task in practice is to estimate the water-use of trees of the current growing season, using models built on data collected in past growing seasons. This section investigates whether the prediction model built on historical data can be used to forecast future sap flux density and water-use without updating the model. To do this, we utilised the sap flux density data collected for the \textit{T. cordata} `18-20' during the 2023 and 2024 growing seasons. Specifically, we built the ensemble prediction model on 2023 data and used it to predict the water-use of 2024 by treating the observed weather data as weather forecast data. During the rolling prediction process, the ensemble elements (i.e., the additive models) were kept unchanged, whereas the ensemble weights were updated each time the prediction window moves forward . 

However, there is one caveat. The scales of the daily cycles of sap flux density vary from tree to tree. From the observed data, they do not appear to be correlated with the stem dimensions. Total tree crown leaf area was not characterised but this is likely to explain some of the variation between trees, even within the same stem diameter class. When historical data are available for the trees to be predicted, they can be integrated (e.g., averaged over to account for the variation or adjusted based on tree physiology) and used to predict the new growing season. When historical data are unavailable, the prediction may be proceeded using models built for trees of the same species. The challenge is to acquire some information about the scale of the daily cycles in order to initialise the rolling prediction procedure. 

\begin{figure}[!htb]
\begin{center}
\includegraphics[width=5.2in]{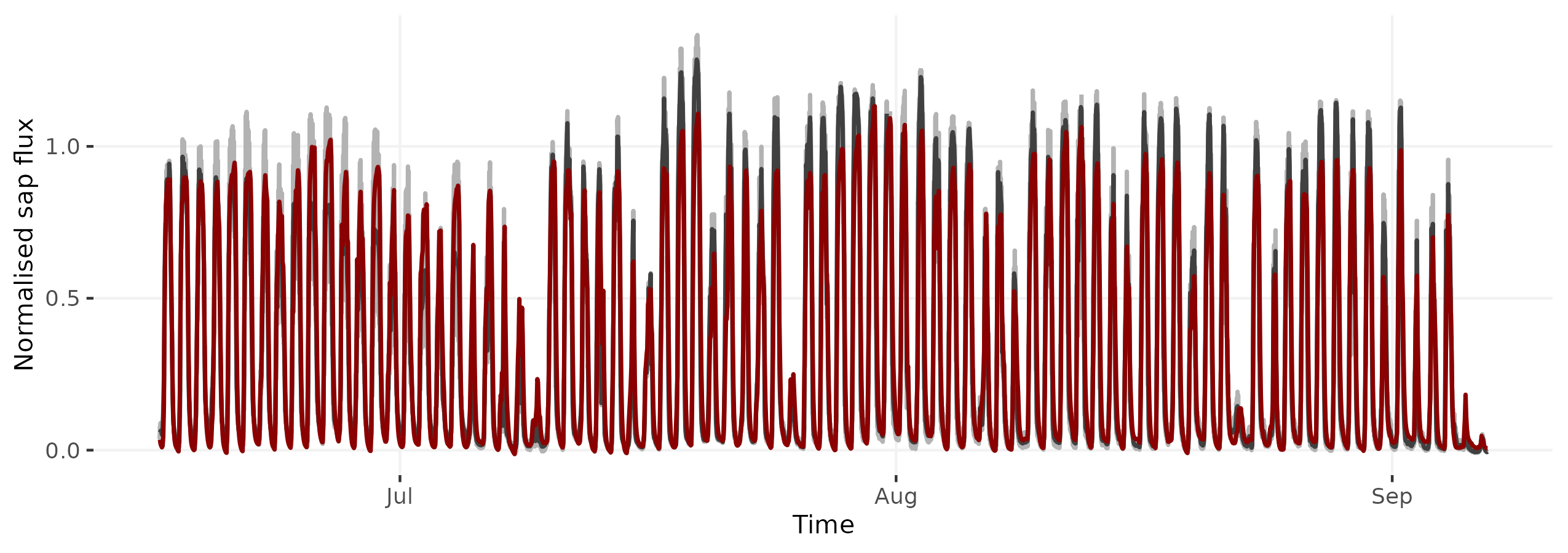}
\includegraphics[width=5.2in]{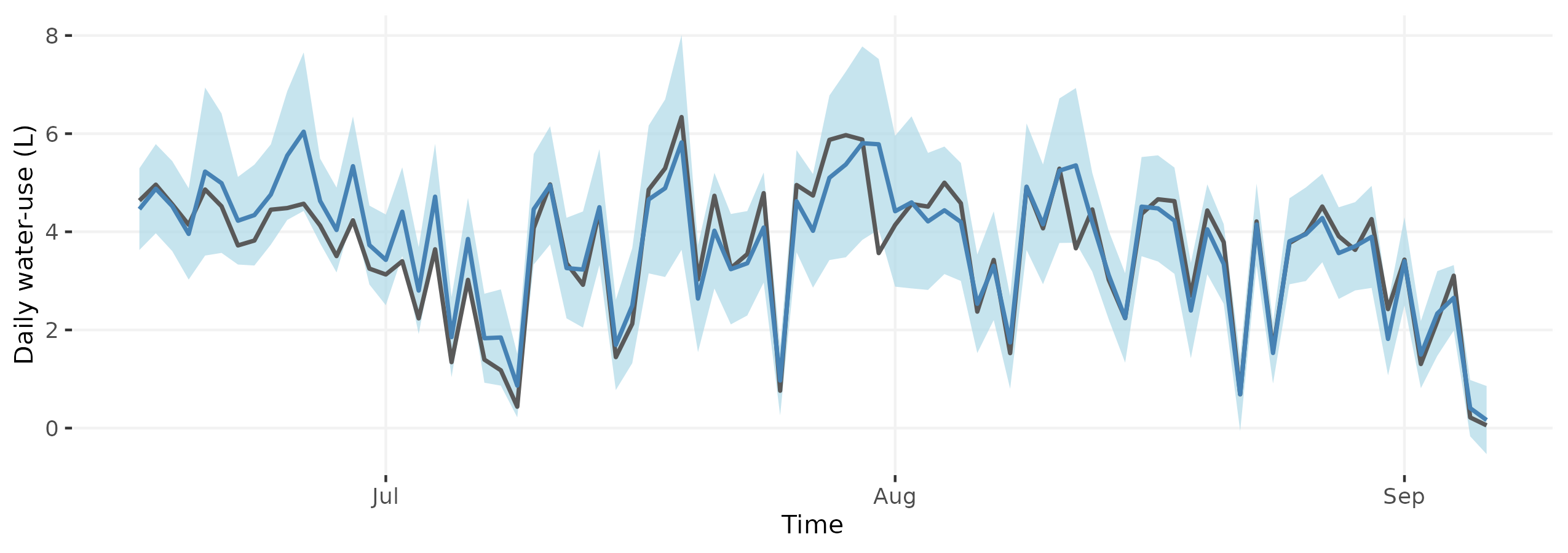}
\captionsetup{labelfont=bf, font=small}
\caption{Normalised hourly sap flux density time series from five \textit{T. cordata} trees (light grey curves) in 2024, the averaged hourly sap flux density (dark grey curve) and the ensemble prediction of sap flux density of a typical tree based on the 2023 model (red curve). (Bottom) Observed daily water-use from the five trees (dark grey curve) in 2024 and the predicted daily water-use from scaling up the ensemble prediction of sap flux density of a typical tree (blue curve), along with uncertainty band of the ensemble prediction (light blue shaded ribbon)}
\label{fig:Tilia_prediction_sapflux}
\end{center}
\end{figure}

Here we present a prediction experiment across two growing seasons on the \textit{T. cordata} group, assuming that certain information about the scale of the sap flux density is known. First the ensemble members (i.e., additive models) were constructed based on the normalised sap flux density data from the past growing season (i.e., 2023 growing season), where normalising was done by re-scaling the sap flux density time series with their 0.95 quantile. Then the ensemble members were used to make rolling ensemble prediction of a normalised sap flux density time series for the 2024 growing season, by treating the 2024 weather data as pseudo weather forecast data. To initialise  the rolling prediction procedure, one may use the sap flux density observations from the same day-of-year in the historical data as the lag-one sap flux density. That is, if the prediction starts on 1 June 2024, then the data from 1 June 2023 can be used to initialise the algorithm. Finally, the predicted normalised time series was scaled back using the `available information' on the scale of daily cycles. In this case, this is taken to be the 0.95 quantile of the observed time series from 2024. Finally, the forecast of daily water-use was produced based on the re-scaled sap flux density prediction. The results were presented in Figure \ref{fig:Tilia_prediction_sapflux}. 

The predicted normalised sap flux density, which is the red curve in the top panel of Figure \ref{fig:Tilia_prediction_sapflux}, follows the pattern of the averaged normalised sap flux density observed in 2024, which is the dark grey curve in the graph. There is some over-estimation in late June and early July, and some under-estimation in late July and August, but the overall performance of the prediction is reasonable. The prediction was then scaled back to estimate the daily water-use of the \textit{T. cordata} `18-20' trees in 2024. Overall, the predicted curve (blue) reflects the pattern in the observed curve (dark grey) well, apart from a small section from the end of July to the beginning of August, where there seem to be larger discrepancy and higher uncertainty (see the bottom panel of Figure \ref{fig:Tilia_prediction_sapflux}). For comparison, another prediction was made using the ensemble model built using the 2024 data. The mean squared errors from the daily water-use prediction based on the 2023 model is 0.2971, and that from the water-use prediction based on the 2024 model is 0.2761. The difference is relatively small.

\section{Discussion} \label{sec:Discussion}

\subsection{The impact of heatwaves on sap flux density}

\begin{figure}[!htb]
\begin{center}
\includegraphics[width=4.6in]{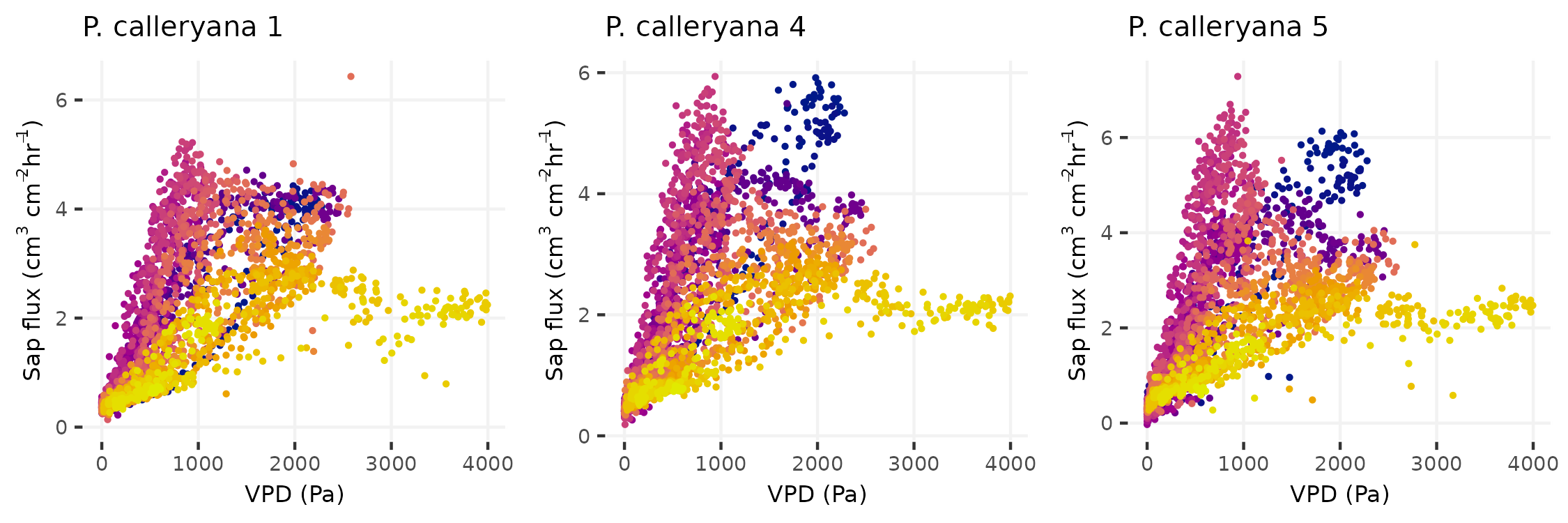}
\includegraphics[width=4.6in]{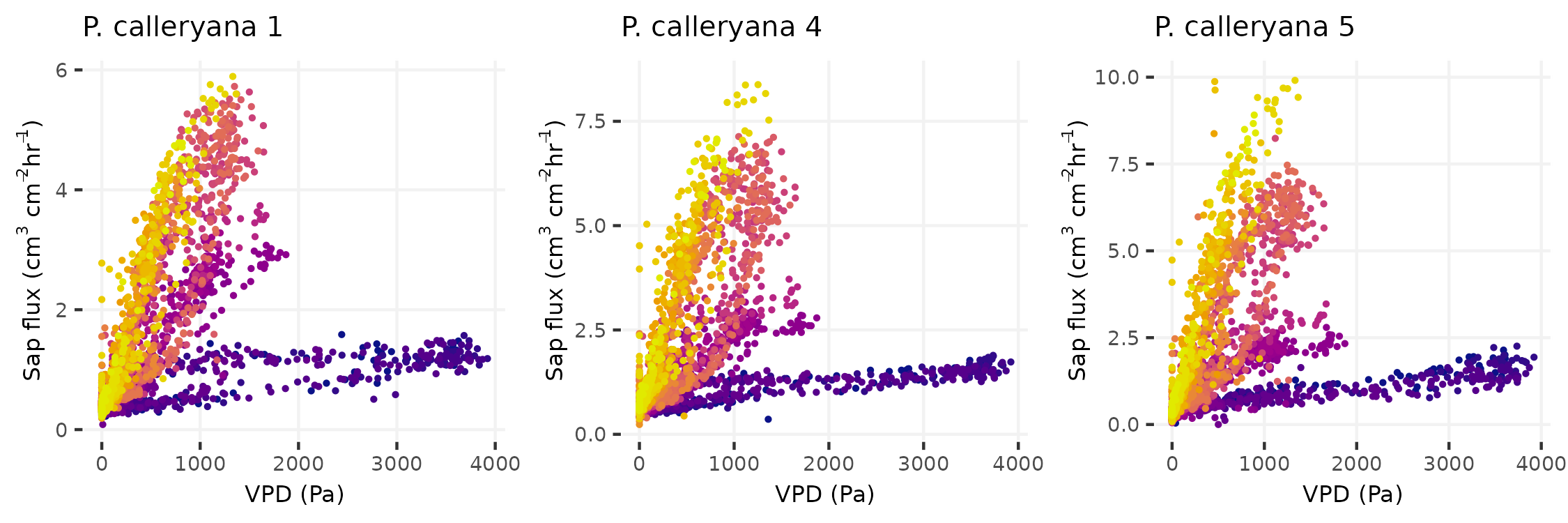}
\captionsetup{labelfont=bf, font=small}
\caption{The scatter plot between sap flux density and VPD for three \textit{P. calleryana} trees, where the points are coloured by calendar day of the period from 16 June to 21 July (top) and from 10 August to 14 September (bottom) in 2022. In all panels, the purple end and the yellow end of the pallette correspond to the beginning and the end of the periods.}
\label{fig:scatter_heatwave_Pyrus}
\end{center}
\end{figure}

The case study in Section \ref{sec:CaseStudy} showed that the proposed ensemble prediction method was able to capture the main patterns in the sap flux density time series and make reasonable predictions of daily water-use for most of the species or groups (see Table \ref{tab:EnsembleExperiment}). The two groups with relatively high prediction errors are \textit{P. calleryana} and \textit{C. betulus} `16-18' in the 2022 growing season, especially the \textit{P. calleryana} group, which has a medium relative prediction error over 30\%. A close inspection showed that the \textit{P. calleryana} trees seemed to have changed their behaviour drastically during and after the two heatwaves in 2022. Heatwaves are known to affect sap flux density and its relationship with the environmental variables \citep{SapfluxHeatwave}. Figure \ref{fig:scatter_heatwave_Pyrus} presents the scatter plots of sap flux density versus VPD, coloured by calendar days from 16 June to 21 July and 10 August to 14 September respectively, for three \textit{P. calleryana} trees. The changes in relationships between the two variables during the two heatwaves, namely from 17 to 19 July (represented by the bright yellow dots in the top panels) and from 9 to 15 August (represented by the dark purple dots in the bottom panels), are clearly visible, indicating potential shifts in the water extraction behaviour as a result of the extreme weather conditions. Using a single model to describe both the normal days and extreme weather days may not be most appropriate, as the relationships between the sap flux density and the environmental variables follow different regimes. 

To investigate the changes in regimes, changepoint analysis was applied to explore the evidence of regime shift and whether the timing of the changes corresponding to the extreme weather events. Changepoint analysis is typically used to identify the timing of abrupt changes in a time series. Here we investigate the residual time series from the additive model (\ref{eqn:GAM_individual}) fitted to the sap flux density data of \textit{P. calleryana} trees from the 2022 growing season. Under the null hypothesis that the model is appropriate for the entire time series, the residuals are expected to follow the same distribution. A significant change in the mean and/or variance of the residual time series can indicate that the global model does not suit certain sections of the time series. As there may be more than one changes in the residual time series, methods for detecting multiple changepoints in mean and variance were considered. Specifically, the CROPS algorithm \citep{PenaltyCpt} with penalised exact linear time \citep{PELT} was used, which can be implemented using functions in the \texttt{changepoint} package in R \citep{changepoint}.  

For investigation purpose, the number of changepoints were chosen to match the changes in the weather pattern, which is four in this case. The results from segmenting the residual time series of \textit{P. calleryana} 1 and 5 with four changepoints suggest some connections between the timings of changes and the heatwave periods. Specifically, the first segment corresponds roughly to the period in early summer when the weather was normal, the second segment corresponds roughly to the period leading to the first heatwave (17 to 19 July 2022), the third segment corresponds to the period from right after the first heatwave till the end of the second heatwave in August (9 to 15 August 2022), the fourth segment corresponds to the short period after the second heatwave, when the trees were trying to recover from the heatwaves, and finally the last segment corresponds to the period when the sap flux density was restored with the increase of soil water availability. Figures showing the changepoint detection results were presented in the supplementary document (Figure S9). 

These results suggest that further improvement of the prediction method may be required, potentially through the use of different models (i.e., models with different parameters, structures or covariates) to capture the changes in tree water extraction behaviour due to heatwaves. In terms of prediction, the main challenge lies in whether it is possible to predict the change of behaviour in trees due to extreme weather events. This is difficult because the effect of extreme weather events on trees may not be immediate. For example, the \textit{P. calleryana} trees appeared to be most affected by the second heatwave, which happened after a long period of relatively high temperature and low precipitation. Their recovery from the heatwave was also gradual and did not fully restore to the pre-heatwave level. In addition, the change of behaviours is not solely determined by the weather. Inter-specific variation in water use varies as a function of stomatal behaviour (e.g., isohydric versus anisohydric) \citep{Stomatal} which will influence the magnitude of sap flux density. Therefore, even with reliable weather forecast data, predicting the change in species-specific tree behaviours can be challenging. Nonetheless, understanding how sap flux density is modulated by extreme weather events interacting with species' hydraulic strategy warrants further investigation as it will substantively benefit the management of tree water-use. 

\subsection{The influence of size class and other environmental factors}

Another factor that can influence the prediction of sap flux density and water-use is the size of trees. It appears that tree water-use does not necessarily scale linearly with the size of the tree. \cite{SizeMatters} noted that sap flow-based transpiration estimates may be particularly sensitive to large trees due to nonlinear relationships between tree-level water use and tree diameter at breast height; \cite{SizeWater} described large variations in the relationship between tree sizes and water-use for individual species or stands, leading to the interesting question of water competition between trees of different sizes. From a practical perspective, tree growers tend to plant trees of similar size and age together for management purpose, and hence the need of estimating the water-use of trees from the same size class. 

In an exploratory analysis on the impact of tree size on sap flux density modelling, additive models with size class as categorical variables were investigated using data collected for \textit{C. betulus} in 2022 and \textit{U.} `New horizon' in 2023 from two different size classes. The estimated linear and non-linear effects from two categories were compared to see if there is any significant difference between the size classes. For the non-linear effect, this refers to whether there is any significant difference in any intervals of the estimated smooth functions. Superior predictive ability (SPA) tests \citep{SPAtest} were used to investigate whether there is a significant difference between the predictions made by the models that used tree size information and models that did not. The results showed some evidence of the impact of size class on the modelling and prediction of sap flux density. Some details on the comparison of the estimated non-linear effect and the implementation of the SPA test to the sap flux density prediction problem are given in the supplementary document. 

Soil water plays an important role in the hydraulic function of trees \cite{SapfluxMaths}. Soil moisture has been identified in literature as another factor that can change the relationship between sap flux density and the environmental variables. For example, \cite{SapfluxAsh} documented a change in relationship between the sap flux density of black ash and VPD from positive to negative as the soil moisture level drops from saturation to a low point; \cite{SapfluxDryForest} noted that different species respond to the decrease of soil moisture differently, with some deciduous tree species reducing sap flux sensitively and some evergreen species increasing their sap flux. It would be interesting to investigate these aspects further to improve the understanding of the response of trees to different internal and external factors and produce more accurate predictions.

\subsection{Potential applications of the proposed method}

Whilst the method proposed in this manuscript has been motivated by the need to estimate water-use of trees grown in a landscape tree nursery, the method itself has broader applications. It can be applied to various different settings in urban forestry, commercial forestry, orchard production, agriculture, and ecosystem restoration. For example, the monitoring and modelling techniques can be implemented on trees in forests to help understand the hydrology of ecosystems. The ensemble prediction method can be applied to provide essential information to water management in horticulture, agriculture and forestry. 

Sap flux monitoring has been recognised as a relatively inexpensive way to obtain an estimate of whole-tree and whole-stand water use of the forest ecosystem \citep{SapfluxSpatial}. For example, the sap flow sensors used in this study, can be installed relatively easily with professional training and fundamental knowledge in tree anatomy. The sensors will then provide high frequency measurements of sap flux density, which gives real-time estimation of the water-use of the monitored trees. By appropriate scaling, the water-use estimation for the entire field/woodland/forest can be obtained efficiently. On the other hand, the proposed additive model for sap flux density, using non-parametric regression, has the potential of capturing various types of relationships between sap flux density and the environmental drivers. Therefore, it can be applied to trees with different hydraulic strategies growing in different climates. It is also possible to replace some of the non-parametric components (i.e., the smooth non-linear relationship) in the model with specific parametric forms, if relevant information is available. Once the update on the additive models is done, the ensemble prediction can be carried out as described in Algorithm \ref{alg:Ensemble} without further change. 

The method can be integrated into real-time monitoring platforms to carry out online prediction of tree water-use. The input of the prediction model is relatively easy to acquire once the sensors are installed and reliable weather forecast data of the field site are available. The computation time to update the ensemble elements and re-train the ensemble weights is minute for daily water-use prediction. Overall, it is a prediction tool that can be easily integrated into any existing monitoring platforms, and be used to assist irrigation planning and decision making.

\section{Conclusion} \label{sec:Conclusion}

This manuscript proposed a novel approach that couples the modern sensor technology and statistical modelling to improve the understanding and prediction of tree sap flux density and daily water-use. The proposed prediction method, based on an ensemble of additive models at its core, has the ability to capture the complex non-linear relationships between sap flux density and its environmental drivers, and to make full use of the available data in producing on-line predictions suitable for assisting irrigation planning. This is crucial to the improvement of irrigation efficiency and to the securing of a sustainable future. The proposed method performed well on real sap flux density time series collected from the field, and has the potential to be integrated into real-time monitoring platform for online prediction, provided that reliable weather forecast data are available. Future improvement of the method may include, (a) accounting for the influence of heatwave and drought in the prediction model via thresholding, and (b) investigating further the influence of tree size for better scaling from the prediction of a standard tree to a group of trees.

% \end{linenumbers}

\vspace{1cm}
\section*{Acknowledgements}

The authors gratefully acknowledge the support from the Tree Production and Innovation Fund (Grant No. TPIF\_52), and the support from Hillier Nurseries for acquiring the tree sap flux density data and weather data during the 2022, 2023 and 2024 growing seasons.  \\

\section*{Author Contributions}

\textbf{Mengyi Gong:} Conceptualization, Methodology, Data analysis, Writing. \\
\textbf{Rebecca Killick:} Conceptualization, Methodology, Writing. \\
\textbf{Andrew Hirons:} Conceptualization, Data collection, Writing. \\

\section*{Conflict of interest statement}

The authors declare that they have no known competing financial interests or personal relationships that could have appeared to influence the work reported in this paper. \\

\section*{Code and data availability}

The R code for implementing the ensemble prediction method will be available on GitHub. The field monitoring data are not publicly available. Please contact the authors for more information. \\


\vspace{1cm}
\begin{thebibliography}{20}

\bibitem[Ahongshangbam \textit{et. al.}(2023)]{SapfluxHeatwave}
Ahongshangbam, J., Kulmala, L., Soininen, J., Fr\"uhauf, Y., Karvinen, E., Salmon, Y., Lintunen, A., Karvonen, A., J\"arvi, L., 2023.
Sap flow and leaf gas exchange response to a drought and heatwave in urban green spaces in a Nordic city.
\textit{Journal of Geophysical Research: Biogeosciences}, 20(21), 4455--4475.
\url{https://bg.copernicus.org/articles/20/4455/2023/}

\bibitem[Asgharinia \textit{et. al.}(2022)]{TreeTalker}
Asgharinia, S., Leberecht, M., Belelli Marchesini, L., Friess, N., Gianelle, D., Nauss, T., Opgenoorth, L., Yates, J., Valentini, R., 2022.
Towards continuous stem water content and sap flux density monitoring: IoT-based solution for detecting changes in stem water dynamics.
\textit{Forests}, 13(7):1040.
\url{https://doi.org/10.3390/f13071040}

\bibitem[Berdanier \textit{et. al.}(2016)]{RadialSapFlux}
Berdanier, A. B.,  Miniat, C. F., Clark, J. S., 2016. 
Predictive models for radial sap flux variation in coniferous, diffuse-porous and ring-porous temperate trees 
\textit{Tree Physiology}, 36, 932--941.

\bibitem[Berry \textit{et. al.}(2018)]{SizeMatters}
Berry, Z. C., Looker, N., Holwerda, F., Aguilar, L. R. G., Colin, P. O., Mart\'inez, T. G., Asbjornsen, H., 2018. 
Why size matters: the interactive influences of tree diameter distribution and sap flow parameters on upscaled transpiration.
\textit{Tree Physiology}, 38 (2), 263--275, \url{https://doi.org/10.1093/treephys/tpx124}

\bibitem[British Standard Institute(1992)]{BS}
British Standard Institute, 1992. BS 3936-1:1992 Nursery stock. \textit{Specification for trees and shrubs}.
London: British Standards Institute. \url{https://doi.org/10.3403/00262241}

\bibitem[Burgess \textit{et. al.}(2001)]{HeatPulseMethod1}
Burgess, S. S., Adams, M. A., Turner, N. C., Beverly, C. R., Ong, C. K., Khan, A. A. Bleby, T. M., 2001.
An improved heat pulse method to measure low and reverse rates of sap flow in woody plants.
\textit{Tree physiology}, 21(9), 589--598.
% \url{https://doi.org/10.3389/fpls.2018.00945}

\bibitem[Butz \textit{et. al.}(2018)]{SapfluxDryForest}
Butz, P., H\"olscher, D., Cueva, E., Graefe, S., 2018.
Tree water use patterns as influenced by phenology in a dry forest of southern Ecuador.
\textit{Frontiers in Plant Science}, 9, 2018.
\url{https://doi.org/10.3389/fpls.2018.00945}

\bibitem[Capezza \textit{et. al.}(2021)]{AdditiveStack}
Capezza, C., Palumbo, Biagio., Goude, Y., Wood, S. N., Fasiolo, M., 2021.
Additive stacking for disaggregate electricity demand forecasting.
\textit{Annals of Applied Statistics}, 15 (2), 727--746.

\bibitem[Claeskens \textit{et. al.}(2016)]{ForecastAvg}
Claeskens, G., Magnus, J. R., Vasnev, A. L., Wang, W., 2016.
The forecast combination puzzle: A simple theoretical explanation.
\textit{International Journal of Forecasting}, 32 (2016), 754--762.

\bibitem[Fahrmeir \textit{et. al.}(2013))]{Regression}
Fahrmeir, L., Kneib, T., Lang, S., Marx, B., 2013.
\textit{Regression: Models, Methods and Applications}.
Springer Berlin, Heidelberg.

\bibitem[Forrester \textit{et. al.}(2022)]{SizeWater}
Forrester, D. I., Limousin, J-M., Pfautsch, S., 2022.
The relationship between tree size and tree water-use: is competition for water size-symmetric or size-asymmetric?
\textit{Tree Physiology}, 42 (10), 1916--1927.
\url{https://doi.org/10.1093/treephys/tpac018}

\bibitem[Gartner \textit{et. al.}(2009)]{SapfluxBirch}
Gartner, K., Nadezhdina, N., Englisch, M., \v Cermak, J., Leitgeb, E., 2009.
Sap flow of birch and Norway spruce during the European heat and drought in summer 2003.
\textit{Forest Ecology and Management}, 258 (5), 590--599.
\url{https://doi.org/10.1016/j.foreco.2009.04.028.}

\bibitem[Hansen(2005)]{SPAtest}
Hansen, P. R., 2005.
A Test for Superior Predictive Ability. 
\textit{Journal of Business \& Economic Statistics}. 
23(4), 365--380.
\url{https://doi.org/10.1198/073500105000000063}

\bibitem[Hastie \textit{et. al.}(2009))]{StatsLearning}
Hastie, T., Tibshirani, R., Friedman, J., 2009.
\textit{The Elements of Statistical Learning: Data Mining, Inference, and Prediction, Second Edition}.
Springer New York, NY.

\bibitem[Haynes \text{et. al.}(2017)]{PenaltyCpt}
Haynes, K., Eckley, I. A., Fearnhead, P., 2017.
Computationally efficient changepoint detection for a range of penalties.
\textit{Journal of Computational and Graphical Statistics}, 126(1), 134--143.

\bibitem[Horma \textit{et. al.}(2011)]{SapfluxPerhumid}
Horna, V., Schuldt, B., Brix, S., Leuschne, C., 2011.
Environment and tree size controlling stem sap flux in a perhumid tropical forest of Central Sulawesi, Indonesia.
\textit{Annals of Forest Science}, 68, 1027--1038.

\bibitem[Kabala \textit{et. al.}(2025)]{MLprediction3}
Kabala, J.P., Massari, C., Niccoli, F., Natali, M., Avanzi, F., Battipaglia, G., 2025.
Reconstruction of the dynamics of sap-flow timeseries of a beech forest using a machine learning approach.
\textit{Agricultural and Forest Meteorology}, 362, 2025, 110379.
\url{https://doi.org/10.1016/j.agrformet.2024.110379.}

\bibitem[Killick \textit{et. al.}(2012)]{PELT}
Killick, R., Fearnhead, P., Eckley, I. A., 2012.
Optimal detection of changepoints with a linear computational cost.
\textit{Journal of the American Statistical Association}, 107(500), 1590--1598.

\bibitem[Killick \textit{et. al.}(2022)]{changepoint}
Killick, R., Haynes, K., Eckley, I. A., (2022). 
changepoint: An R package for changepoint analysis. R package version 2.2.4,
\url{https://CRAN.R-project.org/package=changepoint}

\bibitem[Kim \& Lee(2024)]{NeedleSensor}
Kim, G., Lee, J., 2024.
Micromachined needle-like calorimetric flow sensor for sap flux density measurement in the xylem of plants.
\textit{Scientific Reports}, 14, 14838 (2024). 
\url{https://doi.org/10.1038/s41598-024-65046-9}

\bibitem[Klein(2014)]{Stomatal}
Klein, J., 2014.
The variability of stomatal sensitivity to leaf water potential across tree species indicates a continuum between isohydric and anisohydric behaviours.
\textit{Functional ecology}, 28 (6), 1313--1320. 

\bibitem[Li \textit{et. al.}(2022)]{MLprediction2}
Li, Y., Ye, J., Xu, D., Zhou, G., Feng, H., 2022.
Prediction of sap flow with historical environmental factors based on deep learning technology.
\textit{Computers and Electronics in Agriculture}, 202, 2022, 107400.
\url{https://doi.org/10.1016/j.compag.2022.107400.}

\bibitem[Lens \textit{et. al.}(2013)]{EmbolismResistance}
Lens, F., Tixier, A., Cochard, H., Sperry, J.S., Jansen, S. and Herbette, S., 2013.
Embolism resistance as a key mechanism to understand adaptive plant strategies.
\textit{Current opinion in plant biology}, 16 (3), 287--292.

\bibitem[Leutbecher \& Palmer(2008)]{EnsembleNWP}
Leutbecher, M., Palmer, T. N., 2008.
Ensemble forecasting.
\textit{Journal of Computational Physics}, 227 (2008), 3515--3539.

\bibitem[Ma \textit{et. al.}(2017)]{EnvControlSapflux}
Ma, C., Luo, Y., Shao, M., Li, X., Sun, L., Jia, X., 2017.
Environmental controls on sap flow in black locust forest in Loess Plateau, China.
\textit{Scientific Reports}, 7, 13160.
\url{https://doi.org/10.1038/s41598-017-13532-8}

\bibitem[O'Brien \textit{et. al.}(2004)]{SapfluxTropical}
O'Brien, J. J., Oberbauer, S. F., Clark, D. B., 2004.
Whole tree xylem sap flow responses to multiple environmental variables in a wet tropical forest. 
\textit{Plant, Cell \& Environment}, 27, 551-567
\url{https://doi.org/10.1111/j.1365-3040.2003.01160.x}

\bibitem[Oogathoo \textit{et. al.}(2020)]{SapfluxDrivers}
Oogathoo, S., Houle, D., Duchesne, L., Kneeshaw, D., 2020.
Vapour pressure deficit and solar radiation are the major drivers of transpiration of balsam fir and black spruce tree species in humid boreal regions, even during a short-term drought.
\textit{Agricultural and Forest Meteorology,}, 291, 108063,
\url{https://doi.org/10.1016/j.agrformet.2020.108063}

\bibitem[Schaeybroeck \& Vannitsem(2016)]{EnsembleSpread}
Van Schaeybroeck, B., Vannitsem, S., 2016.
A probabilistic approach to forecast the uncertainty with ensemble spread. 
\textit{Monthly Weather Review}, 144 (1), 451--468.

\bibitem[Shumway \& Stoffer(2017)]{TSAbook}
Shumway, R. H., Stoffer, D. S., 2017. 
\textit{ Time Series Analysis and Its Applications: With R Examples}, 4th Edition, Springer Texts in Statistics, Springer, Cham.

\bibitem[Sperry \& Tyree(1988)]{XylemEmbolism}
Sperry, J. S., Tyree, M. T., 1988.
Mechanism of water stress-induced xylem embolism.
\textit{Plant Physiology}, 88 (3), 581--587.

\bibitem[Steppe \textit{et. al.}(2006)]{SapfluxMaths}
Steppe, K., De Pauw, D. J. W., Lemeur, R., Vanrolleghem, A., 2006.
A mathematical model linking tree sap flow dynamics to daily stem diameter fluctuations and radial stem growth
\textit{Tree Physiology}, 26 (3), 257--273. \url{https://doi.org/10.1093/treephys/26.3.257}

\bibitem[Stone \& Bel(2022)]{SapfluxCherry}
Stone, C. H., Close, D. C., Corkrey, R., Goodwin, I., 2022.
Sap flow of sweet cherry reveals distinct effects of humidity and wind under rain covered and netted protected cropping systems.
\textit{Scientific Reports}, 12, 21031 (2022).
\url{https://doi.org/10.1038/s41598-022-25207-0}

\bibitem[Strobach \& Bel(2017)]{EnsembleError}
Strobach, E., Bel, G., 2017.
Quantifying the uncertainties in an ensemble of decadal climate predictions. 
\textit{Journal of Geophysical Research: Atmospheres}, 122 (24), 13191--13200.

\bibitem[Su\'arez \textit{et. al.}(2021)]{SapfluxRadiation}
Su\'arez, J. C., Casanoves, F., Bieng, M. A. N., Melgarejo, L. M., Di Rienzo, J. A., Armas, C., 2021
Prediction model for sap flow in cacao trees under different radiation intensities in the western Colombian Amazon.
\textit{Scientific Reports}, 10512 (2021).
\url{https://doi.org/10.1038/s41598-021-89876-z}

\bibitem[Telander \textit{et. al.}(2015)]{SapfluxAsh}
Telander, A. C., Slesak, R. A., D’Amato, A. W., Palik, B. J., Brooks, K. N., Lenhart, C. F., 2015.
Sap flow of black ash in wetland forests of northern Minnesota, USA: Hydrologic implications of tree mortality due to emerald ash borer.
\textit{Agricultural and Forest Meteorology}, 206, 4--11.
\url{https://doi.org/10.1016/j.agrformet.2015.02.019}

\bibitem[Tu \textit{et. al.}(2019)]{MLprediction1}
Tu, J., Wei, X., Huang, B., Fan, H., Jian, M., Li, W., 2019
Improvement of sap flow estimation by including phenological index and time-lag effect in back-propagation neural network models.
\textit{Agricultural and Forest Meteorology}, 276–277, 2019, 107608,
\url{https://doi.org/10.1016/j.agrformet.2019.06.007.}

\bibitem[Vandegehuchte \& Steppe(2013)]{HeatPulseMethod2}
Vandegehuchte, M., Steppe, K., 2013.
Sap-flux density measurement methods: working principles and applicability.
\textit{Functional Plant Biology}, 40(3), 213--223. 

\bibitem[Van de Wal \textit{et. al.}(2015)]{SapfluxSpatial}
Van de Wal, B.A.E., Guyot, A., Lovelock, C.E. Lockington, D. A., Steppe, K., 2015.
Influence of temporospatial variation in sap flux density on estimates of whole-tree water use in \textit{Avicennia marina}. 
\textit{Trees}, 29, 215–-222. \url{https://doi.org/10.1007/s00468-014-1105-z}

\bibitem[Wan \textit{et. al.}(2023)]{LagDependence1}
Wan, L., Zhang, Q., Cheng, L., Liu, Y., Qin, S., Xu, J., Wang, Y., 2023.
What determines the time lags of sap flux with solar radiation and vapor pressure deficit?
\textit{Agricultural and Forest Meteorology}, 333 (109414).
\url{https://doi.org/10.1016/j.agrformet.2023.109414.}

\bibitem[Wan \textit{et. al.}(2024)]{LagDependence2}
Wan, L., Zhang, Q., Arain, M. A., and Cheng, L., 2024.
A novel crossed hysteresis response pattern of sap flux to solar radiation.
\textit{Journal of Geophysical Research: Biogeosciences}, 129, e2024JG007998.
\url{https://doi.org/10.1029/2024JG007998}

\bibitem[Wood(2001)]{mgcv}
Wood, S. N., 2001.
Fast stable restricted maximum likelihood and marginal likelihood estimation of semiparametric generalized linear models.
\textit{Journal of the Royal Statistical Society (B)}, 73 (1), 3--36.

\bibitem[Wood(2017)]{GAM}
Wood, S. N., 2017.
\textit{Generalized Additive Models: An Introduction with R, Second Edition}.
CRC Press, United Kingdom.

\end{thebibliography}
\end{document}

\title{\textbf{Supplementary document to the manuscript ``An ensemble prediction method for forecasting sap flux density and water-use in temperate trees''}}

\author{Mengyi Gong, Rebecca Killick, Andrew Hirons}

\date{}
% \date{\today}

\maketitle

\setstretch{1}

\section{Prediction model for sap flux density}

This section provides additional details to the development of the prediction model for sap flux density. These include, (1) whether it is necessary to consider lagged covariates in the prediction model, and (2) examples of how the models perform in predicting the hourly sap flux density and daily water-use for an individual tree or a group of trees. 

\subsection{Lagged dependence and why it is not considered in prediction models}

The lagged dependence between sap flux density and the environmental drivers, such as VPD and air temperature, is commonly found in many species of trees. Here we investigated the lagged dependence between hourly sap flux density, VPD, solar radiation, air temperature and humidity and whether this information can improve the prediction of sap flux density, using the data collected for \textit{Betula pendula} and \textit{Tilia cordata} during the 2022 and 2024 growing seasons. In particular, we used cross correlations and hysteresis graphs to investigate the existence of the lagged dependence. We then compare the additive models built using lagged covariates to the models built using covariates from the current time in terms of their prediction performance. 

The cross correlation $Corr(X_{t+\tau}, Y_{t})$ measures the correlation between two variables $X$ and $Y$ with a time lag $\tau$. A strong correlation between $X_{t+\tau}$ and $Y_{t}$ for a $\tau > 0$ may be an indicator of a leading effect, of $X$ on $Y$. The hysteresis graph is another way of investigating lagged dependence. The graph typically plots the response variable (on y-axis) against the input/driving variable (on x-axis). The shape and direction (clockwise or counter-clockwise) of the hysteresis loop can provide indications of the leading/lagging relationship between variables. The hysteresis response of tree sap flux to its driving factors has been reported in \cite{LagDependence1} and \cite{LagDependence2}. 

\begin{figure}[!htb]
\begin{center}
\includegraphics[width=4in]{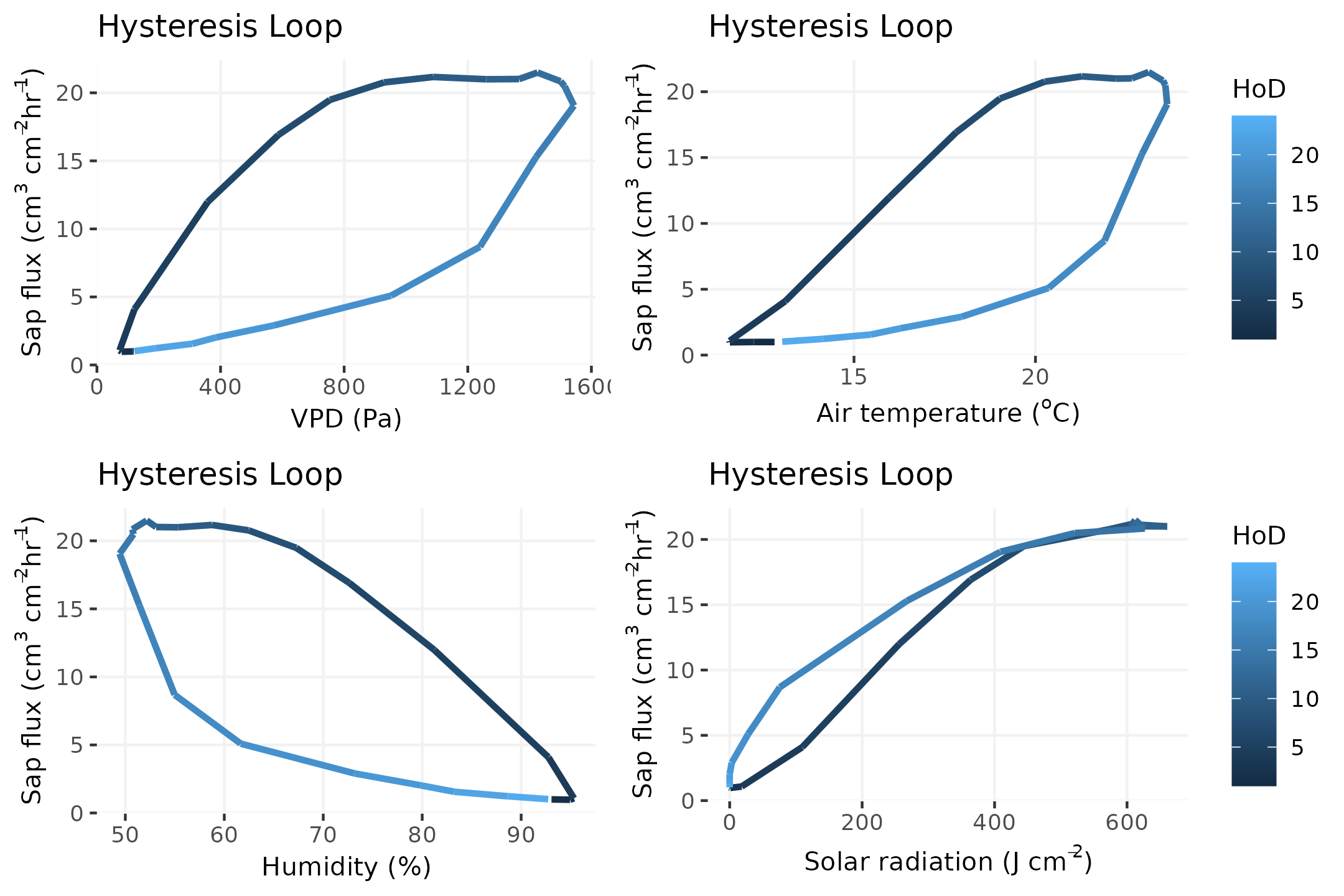}
\captionsetup{labelfont=bf, font=small}
\caption{Hysteresis graphs showing the lagged dependence between sap flux density and the four weather variables (VPD, air temperature, humidity and solar radiation) for \textit{B. pendula} 4 during the period ofs 1 to 15 July 2022. }
\label{fig:Hysteresis_Betula4}
\end{center}
\end{figure}

\begin{figure}[!htb]
\begin{center}
\includegraphics[width=4in]{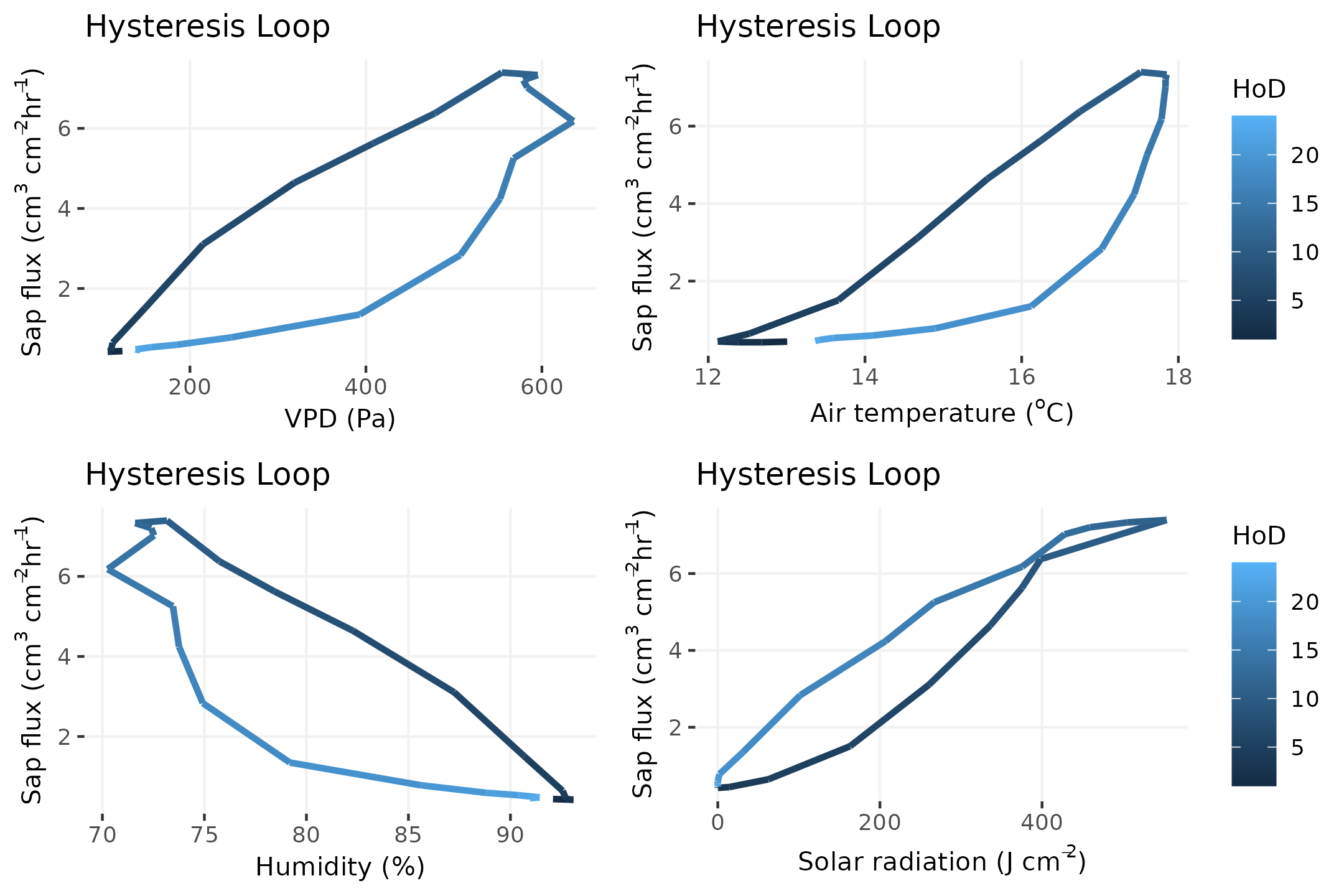}
\captionsetup{labelfont=bf, font=small}
\caption{Hysteresis graphs showing the lagged dependence between sap flux density and the four weather variables (VPD, air temperature, humidity and solar radiation) for \textit{T. cordata} 1 during the period of 1 to 15 July 2024. }
\label{fig:Hysteresis_Tilia1}
\end{center}
\end{figure}

In this case, we see leading effect of solar radiation to sap flux density for the \textit{B. pendula} and \textit{T. cordata} trees, with the largest cross correlations appear at lag-one for majority of the cases. This appears to be consistent with the observations in literature. The hysteresis graphs between sap flux density and the four weather variables for \textit{B. pendula} 4 between 1 and 15 July 2022 and for \textit{T. cordata} 1 between 1 and 15 July 2024 are presented in Figures \ref{fig:Hysteresis_Betula4} and \ref{fig:Hysteresis_Tilia1}. These graphs do not display the typical shape of the hysteresis loop. However, for most of the variables, the lagged or leading effect can be seen from at least part of the hysteresis loops. For example, in the graphs of sap flux density against solar radiation, the lagged effect of sap flux density can be identified in the early morning hours and the hours right after mid-day. Based on these results, we move on to compare two different additive models,
\begin{equation}
Y_{t} = \alpha_{0} + \alpha_{1} Y_{t-1} + \alpha_{2} T_{t} + \alpha_{3} H_{t} + \bm{s}_{1}(R_{t})^{\top} \bm{\beta}_{1} + ( \bm{s}_{2}(V_{t}) \cdot R_{t} )^{\top} \bm{\beta}_{2} + \bm{s}_{3}(V_{t} , X_{t})^{\top} \bm{\beta}_{3} + \epsilon_{t} \; ,
\label{eqn:GAM_individual}
\end{equation}
which uses all current-time covariates, and 
\begin{equation}
Y_{t} = \alpha_{0} + \alpha_{1} Y_{t-1} + \alpha_{2} T_{t} + \alpha_{3} H_{t} + \bm{s}_{1}(R_{t-1})^{\top} \bm{\beta}_{1} + ( \bm{s}_{2}(V_{t}) \cdot R_{t-1} )^{\top} \bm{\beta}_{2} + \bm{s}_{3}(V_{t} , X_{t})^{\top} \bm{\beta}_{3} + \epsilon_{t} \; ,
\label{eqn:GAM_individual_lag}
\end{equation}
which uses the lag-one dependence between sap flux density and solar radiation. Note that the scatterplot between solar radiation and lag-one sap flux density also show non-linear relationships. Hence we continued using the smooth non-linear term $\bm{s}_{1}(R_{t-1})^{\top} \bm{\beta}_{1}$.  

\begin{figure}[!htb]
\begin{center}
\includegraphics[width=5in]{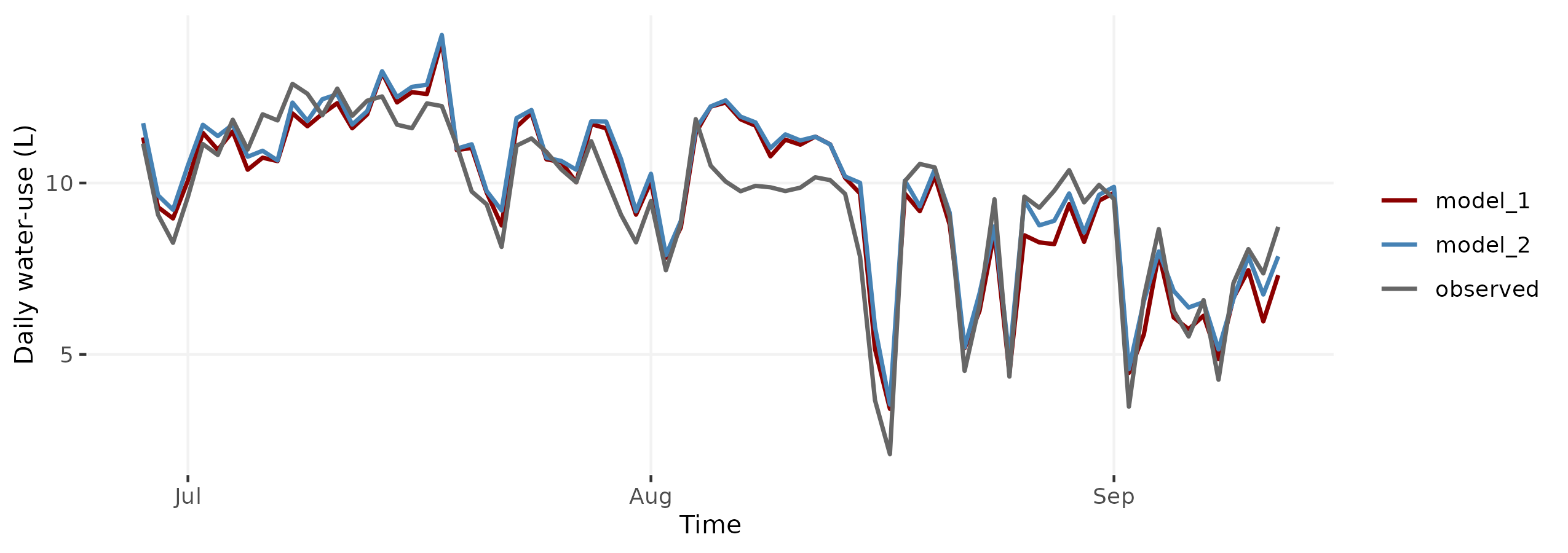}
\includegraphics[width=5in]{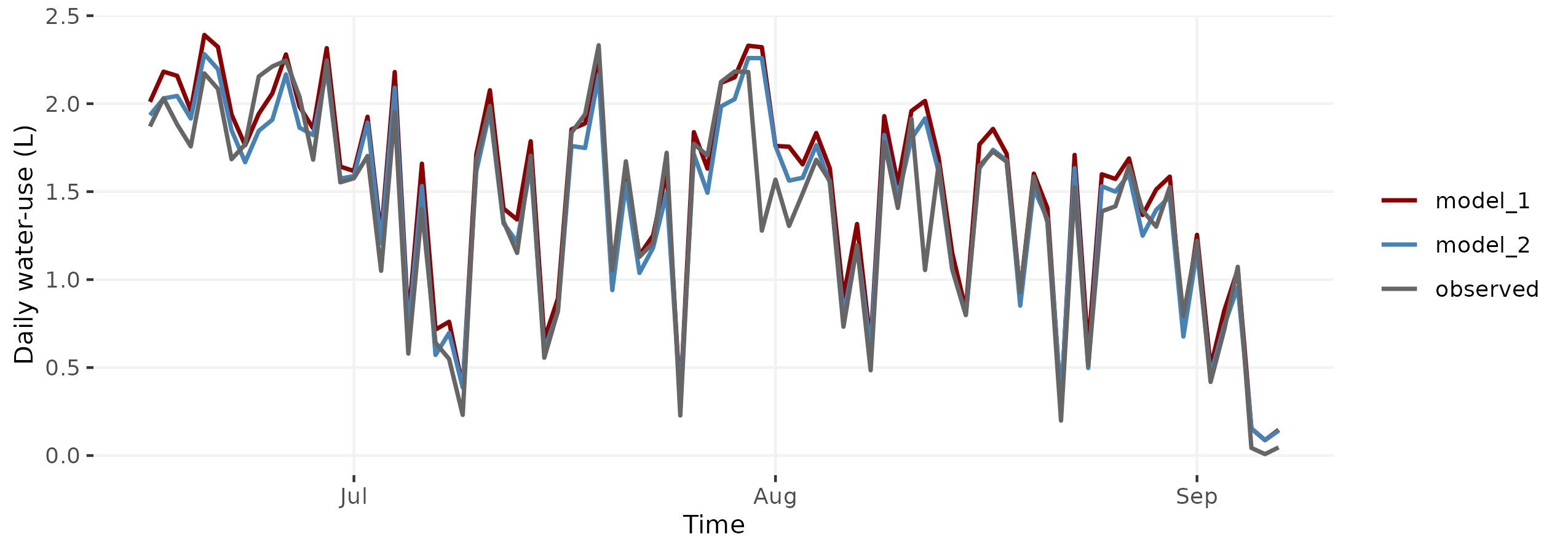}
\captionsetup{labelfont=bf, font=small}
\caption{Observed (grey curves) and predicted daily water-use time series based on model (1) (red curves) and model (2) (blue curves) for \textit{B. pendula} 4 (top) and \textit{T. cordata} 1 (bottom). }
\label{fig:Lag_prediction}
\end{center}
\end{figure}

We investigated both the model fitting and model prediction performance (via mean squared prediction error, or MSPE) of the two models, applied to different \textit{B. pendula} and \textit{T. cordata} trees, using the \texttt{mgcv} package \citep{mgcv} in R. Overall, the models using current-time covariates achieved a better fit (in terms of deviance explained and residual sum of squares) and a better prediction result (in terms of MSPE). Figure \ref{fig:Lag_prediction} presents two examples of the prediction of daily water-use based on the predicted sap flux density from model (1) and (2) using a rolling window prediction approach similar to the one proposed in the main manuscript (simply without the ensemble). We see that the two models perform similarly, with the models using lag-one covariates having slightly larger prediction errors. Therefore, we chose to consider only the current-time covariates for the prediction task in the main manuscript. This can save the time spent on selecting the appropriate order of the lagged dependence, without compromising the prediction performance. For the investigation or interpretation of the nature of the relationships between sap flux density and its drivers, it is important to identify the correct lagged dependence.

\subsection{Examples of predicting sap flux density of individual trees}

Here we present a few examples of predicting the sap flux density of individual trees using the generic additive model introduced in section 2.2 of the main manuscript. In the first example, we built additive models for \textit{B. pendula} 1 from the 2022 growing season. In particular, we investigated models with flexible covariate $X_{t}$ being daily maximum temperature, daily minimum humidity and daily mean soil moisture respectively. It appears that the model using daily maximum temperature best describes the fluctuation in daily scales for \textit{B. pendula} 1. Predictions of sap flux density were then produced using the selected models respectively. Daily water-use prediction were calculated by multiplying the predicted sap flux density with the corresponding sapwood area and then aggregate over the day. The results are presented in Figure \ref{fig:gam_Betula1} below. We applied the same method to \textit{T. cordata} 4 from the 2024 growing season. In this case, the model using daily mean soil moisture best describes the fluctuation in daily scales. The results are presented in Figures \ref{fig:gam_Tilia4} below.

\begin{figure}[!htb]
\begin{center}
\includegraphics[width=5.2in]{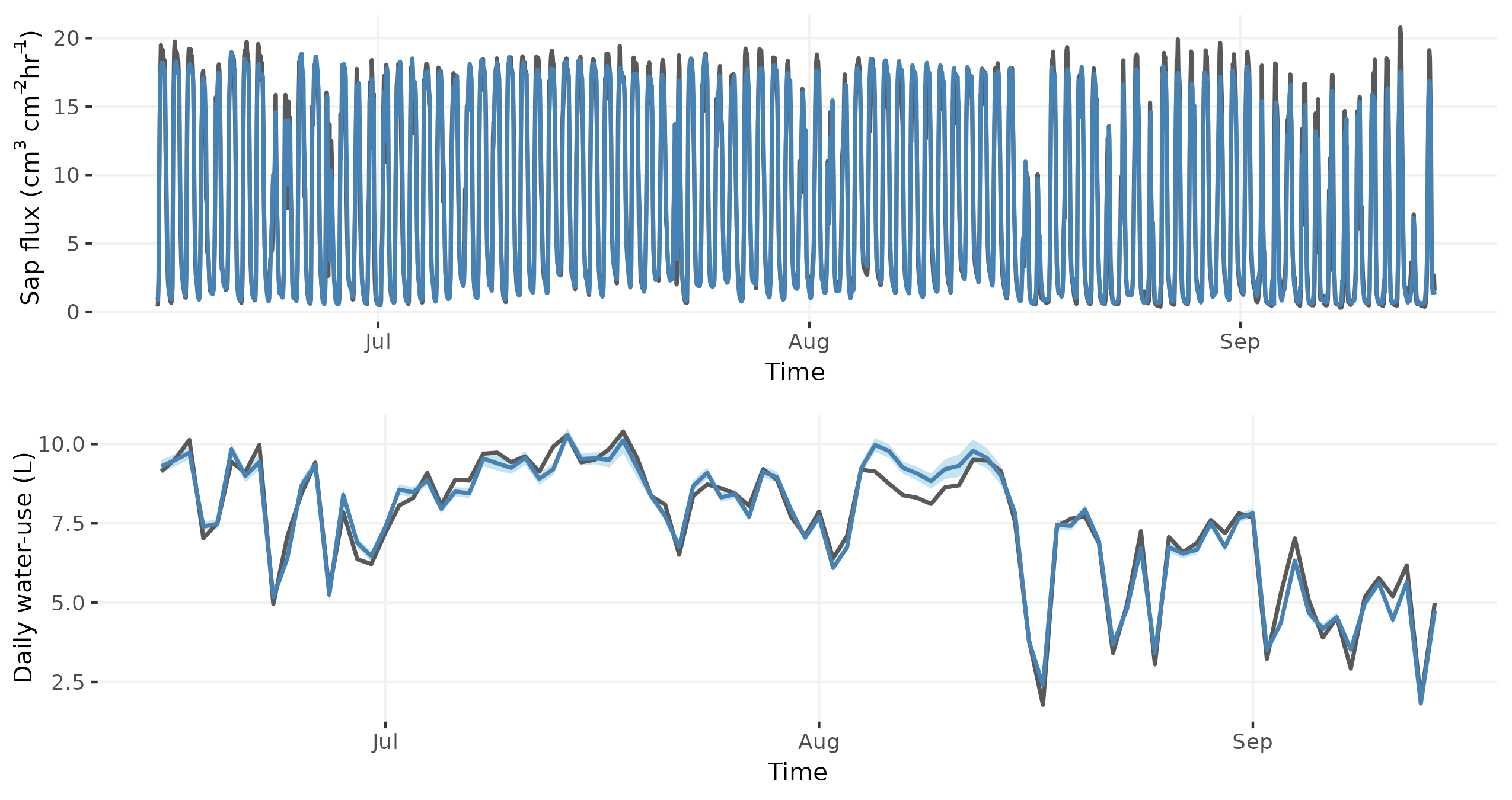}
\captionsetup{labelfont=bf, font=small}
\caption{(Top) Observed (grey) and predicted (blue) hourly sap flux density time series for \textit{B. pendula} 1, using daily maximum temperature as the flexible covariate to explain the variation in daily scales. (Bottom) Observed (grey) and predicted (blue) daily water-use with uncertainty band (light blue shaded band). }
\label{fig:gam_Betula1}
\end{center}
\end{figure}

\begin{figure}[!htb]
\begin{center}
\includegraphics[width=5.2in]{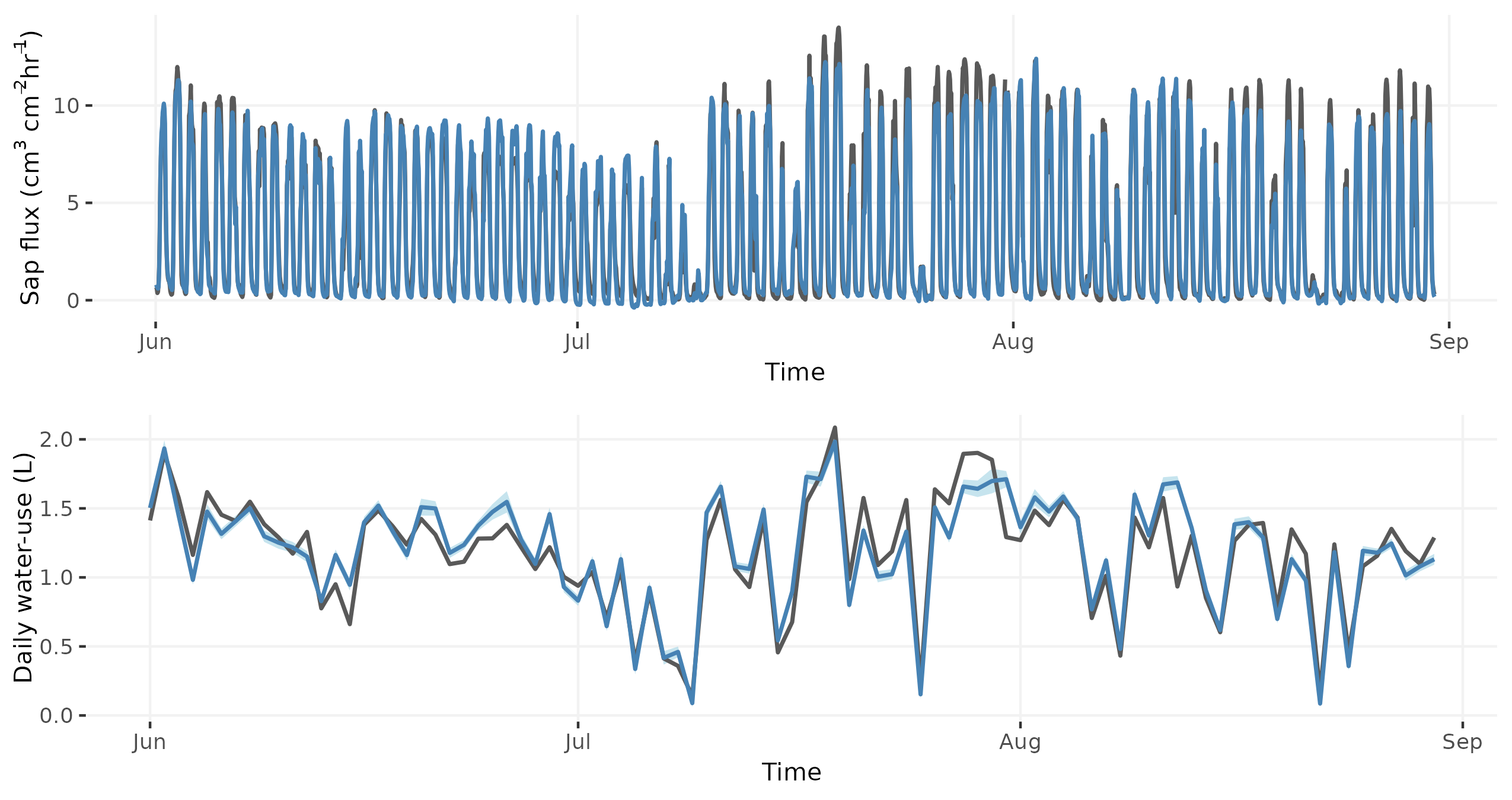}
\captionsetup{labelfont=bf, font=small}
\caption{(Top) Observed (grey) and predicted (blue) hourly sap flux density time series for \textit{T. cordata} 4, using daily mean soil moisture as the flexible covariate to explain the variation in daily scales. (Bottom) Observed (grey) and predicted (blue) daily water-use with uncertainty band (light blue shaded band). }
\label{fig:gam_Tilia4}
\end{center}
\end{figure}

\subsection{Examples of predicting daily water-use of a group of trees}

Here we present examples of rolling ensemble prediction of daily water-use for a group of trees. Two examples, one based on \textit{Tilia cordata} in the 2023 growing season and one based on \textit{Pyrus calleryana} in the 2022 growing season are considered. The two examples represent a reasonable and a problematic prediction respectively. For both species, the ensemble elements (i.e., the additive models of individual trees) were updated based on the pseudo `real-time monitoring' data collected up to the `current' time point, the ensemble weights were retrained based on the last two-weeks' data, and then the hourly sap flux density and the daily water-use prediction for the next prediction window (i.e., the next seven days in this case). 

\begin{figure}[!htb]
\begin{center}
\includegraphics[width=5.2in]{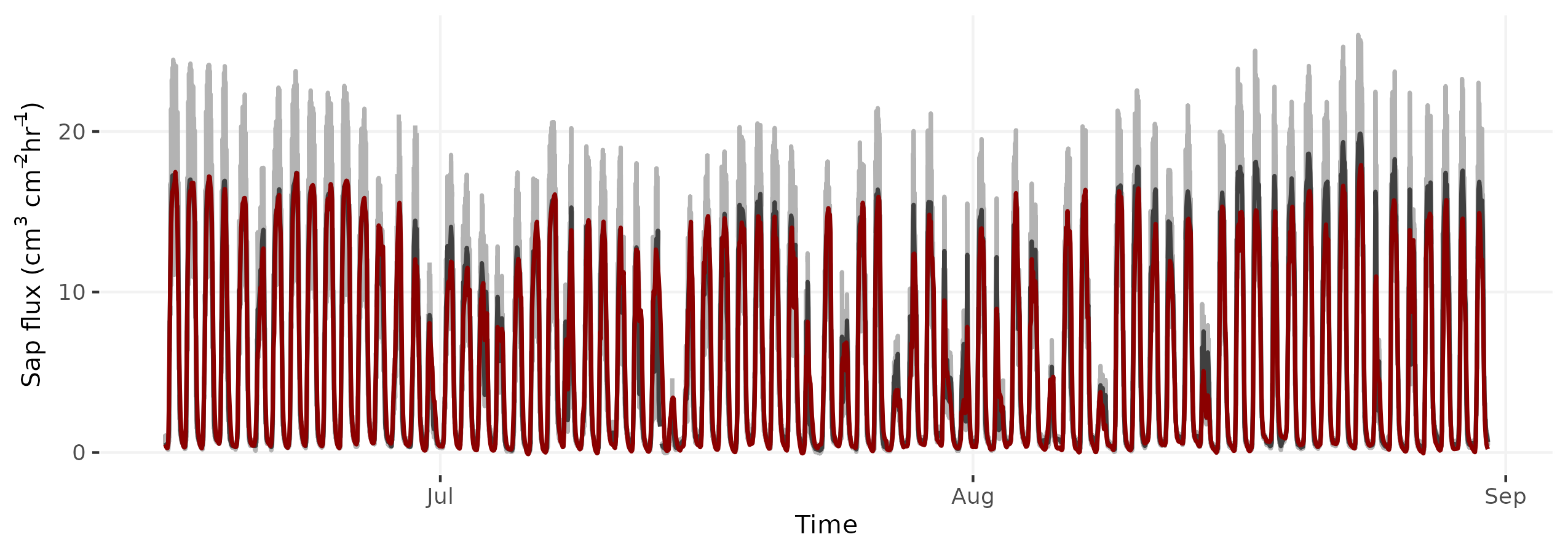}
\includegraphics[width=5.2in]{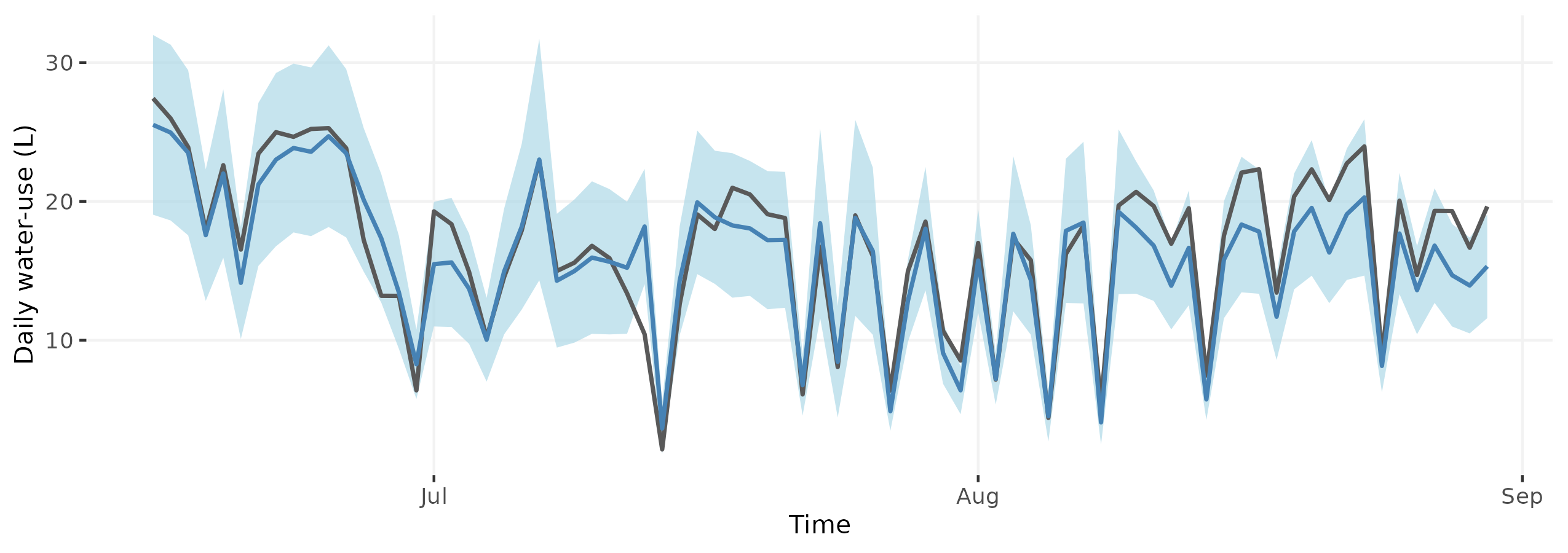}
\captionsetup{labelfont=bf, font=small}
\caption{(Top) Hourly sap flux density time series from five \textit{T. cordata} trees (light grey curves) in 2023, the averaged hourly sap flux density (dark grey curve) and the ensemble prediction of sap flux density of a typical tree (red curve). (Bottom) Observed daily water-use from five trees (dark grey curve) in 2023 and the predicted daily water-use from scaling up the ensemble prediction of sap flux density of a typical tree (blue curve), along with uncertainty band of the ensemble prediction (light blue shaded ribbon).}
\label{fig:Tilia_ensemble}
\end{center}
\end{figure}

Figure \ref{fig:Tilia_ensemble} presents an example of the ensemble prediction of daily water-use of \textit{T. cordata}, which resulted in a good prediction performance. The top panel of Figure \ref{fig:Tilia_ensemble} shows the prediction of the sap flux density time series of a standard \textit{T. cordata} tree using models built on the data of five `18-20' \textit{T. cordata} trees from the 2023 growing season. In this case, the prediction curve follows the observed curve majority of the time, apart from the second half of August, where there seems to be an under-estimate of water-use (see bottom panel of Figure \ref{fig:Tilia_ensemble}). Nonetheless, the predicted water-use time series falls within the uncertainty band, which was obtained by scaling the prediction uncertainty estimated from the ensemble spread as described in the manuscript. 

\begin{figure}[!htb]
\begin{center}
\includegraphics[width=5.2in]{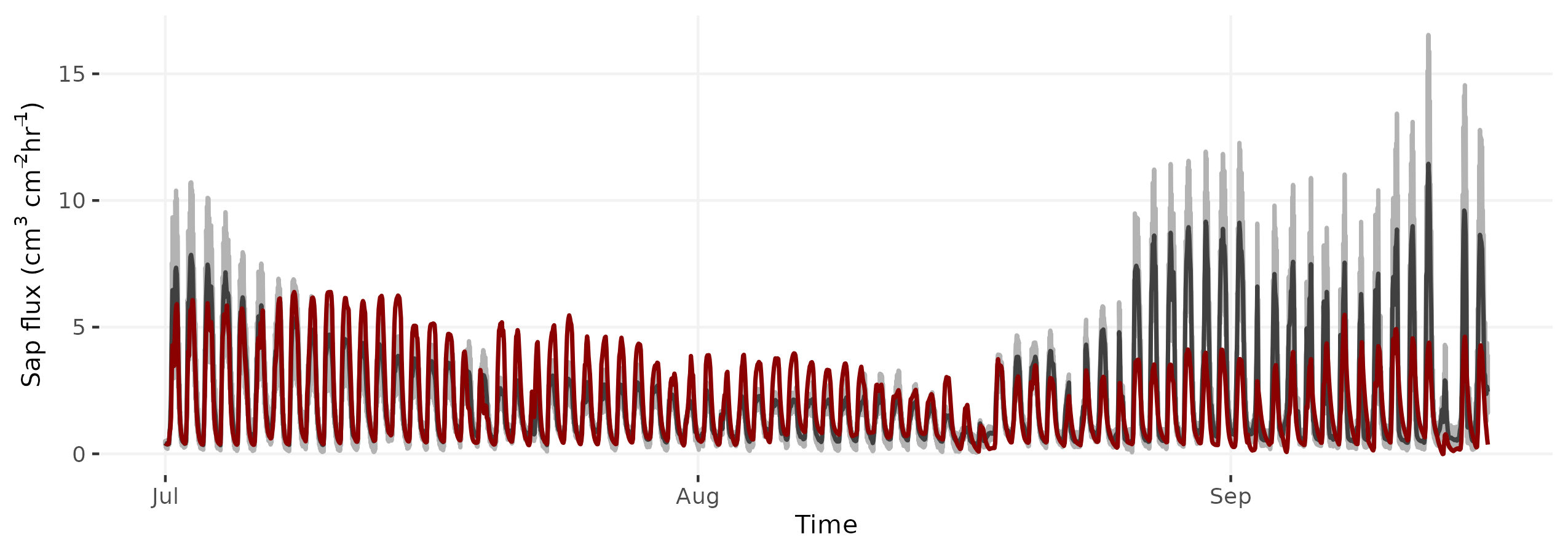}
\includegraphics[width=5.2in]{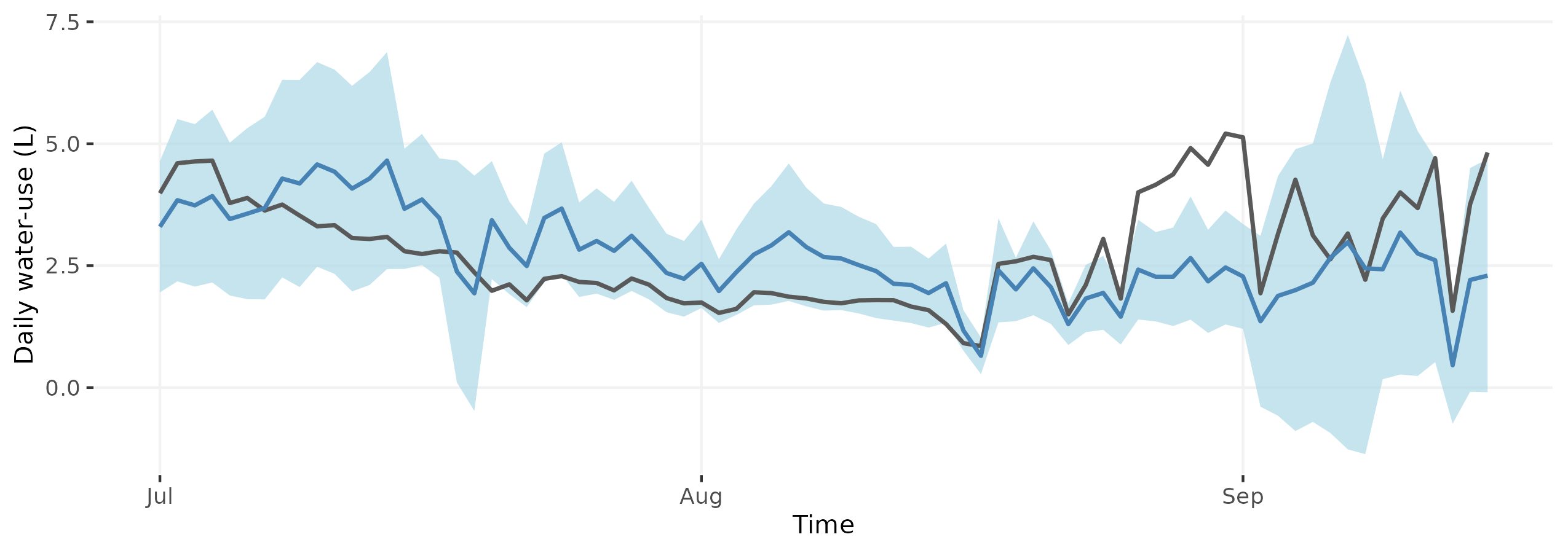}
\captionsetup{labelfont=bf, font=small}
\caption{(Top) Hourly sap flux density time series from three \textit{P. calleryana} trees (light grey curves) in 2022, the averaged hourly sap flux density (dark grey curve) and the ensemble prediction of sap flux density of a typical tree (red curve). (Bottom) Observed daily water-use from three trees (dark grey curve) in 2022 and the predicted daily water-use from scaling up the ensemble prediction of sap flux density of a typical tree (blue curve), along with uncertainty band of the ensemble prediction (light blue shaded ribbon).}
\label{fig:Pyrus_ensemble}
\end{center}
\end{figure}

Figure \ref{fig:Pyrus_ensemble} presents an example of the ensemble prediction of daily water-use of \textit{P. calleryana}, which resulted in a problematic prediction, potentially due to the impact from the heatwaves in summer 2022. The top panel of Figure \ref{fig:Pyrus_ensemble} shows the prediction of the sap flux density time series of a standard tree using models built on the data of three \textit{P. calleryana} trees from the 2022 growing season. In this case, the predicted time series over-estimates the observed sap flux time series in July and the first half of August, and then significantly under-estimates the observed time series from the second half of August onward. The 95\% prediction interval of daily water-use does not cover the actual water-use for a period at the end of August (see bottom panel of Figure \ref{fig:Pyrus_ensemble}). This example suggests that the ensemble elements struggle to capture the temporal patterns in the data, and that further improvement of the modelling approach is needed.

\section{Investigating the impact of heatwaves}

This section provides some more details on the impact of heatwaves on the sap flux prediction models. For this investigation, we will use the data from the \textit{P. calleryana} trees from the 2022 growing season. As shown in Figure \ref{fig:Pyrus_ensemble}, the prediction model performed poorly on the \textit{P. calleryana} trees. We suspect this is related to the impact of heatwaves. 

\subsection{Temporal patterns in sap flux density during heatwaves}

\begin{figure}[!htb]
\begin{center}
\includegraphics[width=5in]{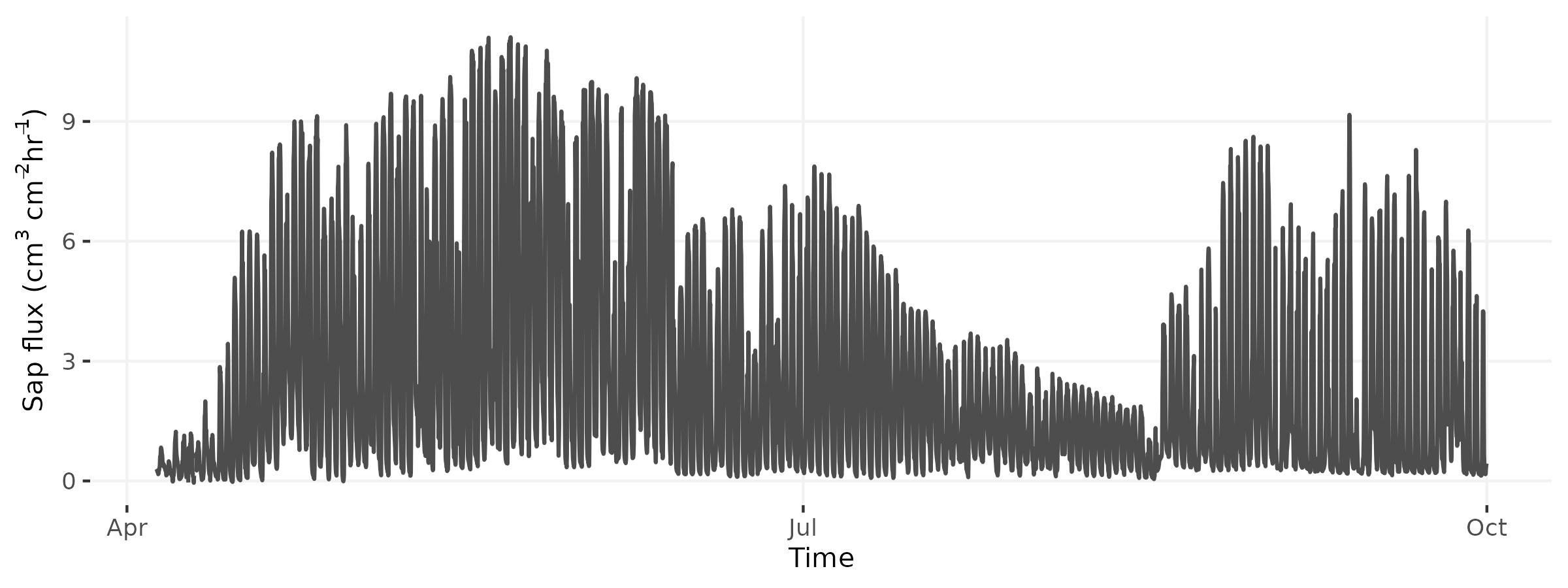}
\includegraphics[width=5in]{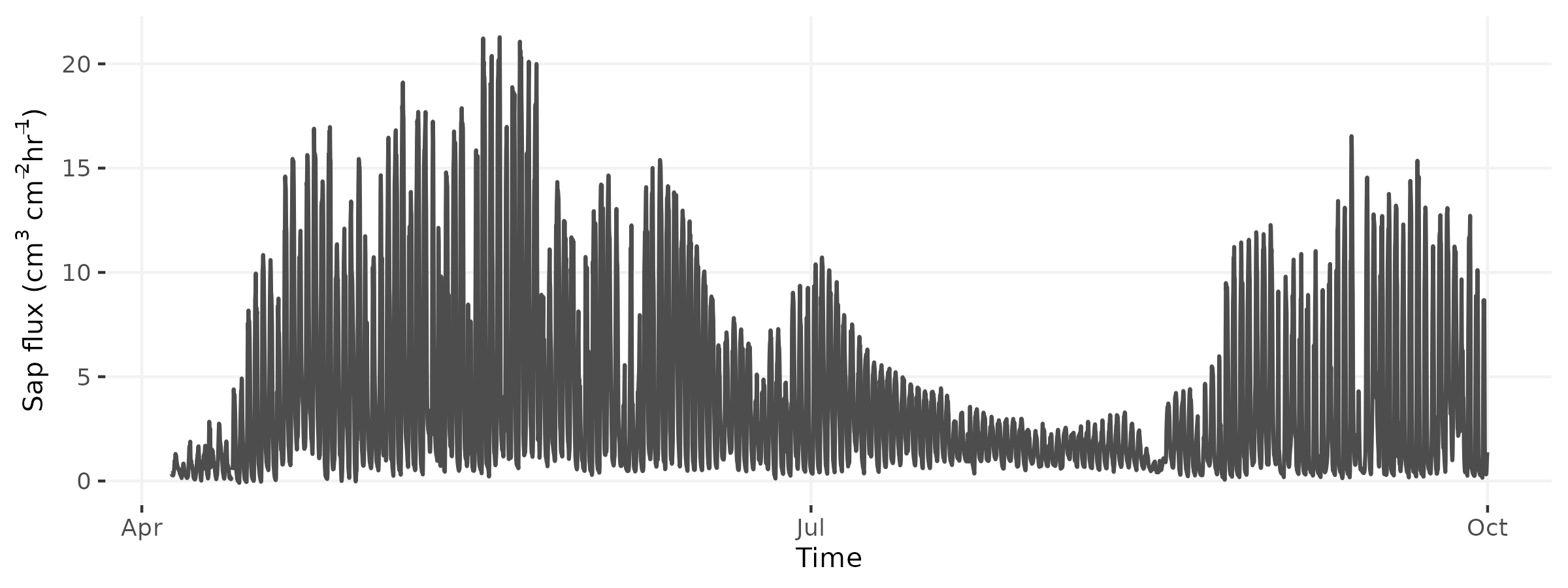}
\captionsetup{labelfont=bf, font=small}
\caption{Sap flux density time series of \textit{P. calleryana} 1 (top) and \textit{P. calleryana} 5 (bottom) during the 2022 growing season.}
\label{fig:PyrusStress}
\end{center}
\end{figure}

Figure \ref{fig:PyrusStress} shows the sap flux density time series of \textit{P. calleryana} 1 and \textit{P. calleryana} 5 from the beginning of April to the end of August 2022. The time series at the beginning of the growing season followed similar pattern as many other species, with the daily peaks of sap flux density increasing as the leaves expanding, until it reached a high level. Then the daily peaks started to shrink as the temperature getting higher and higher from mid June onward. This was followed by a more dramatic drop in daily peaks when the two heatwaves hit in late July and mid August. The daily peaks during the second heatwave were lower than 1/3 of the daily peaks in June for \textit{P. calleryana} 1 and lower than 1/5 of the daily peaks in June for \textit{P. calleryana} 5. After the second heatwave when the water availability was restored in late August, the daily peaks of sap flux density were restored to some extent. However, neither of the trees recovered to their pre-heatwave levels. This highlights the significant impact of prolonged high temperature and lack of precipitation have on trees. As discussed at the end of section 3.1 of the main manuscript, the decline in soil water availability during a period of high VPD days can result in water potential that induces embolism in the xylem vessels, leading to the concomitant reduction of hydraulic conductivity, depressing the sap flux density.

\subsection{Changepoint analysis on model residuals}

\begin{figure}[!htb]
\begin{center}
\includegraphics[width=5in]{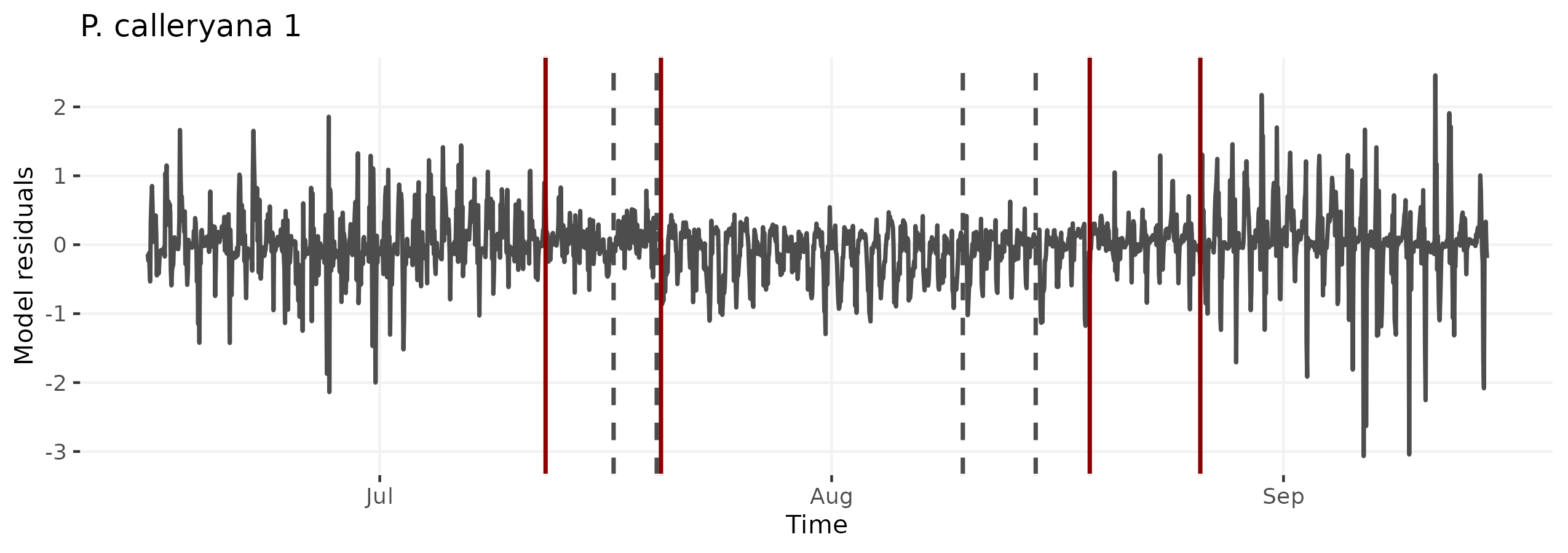}
\includegraphics[width=5in]{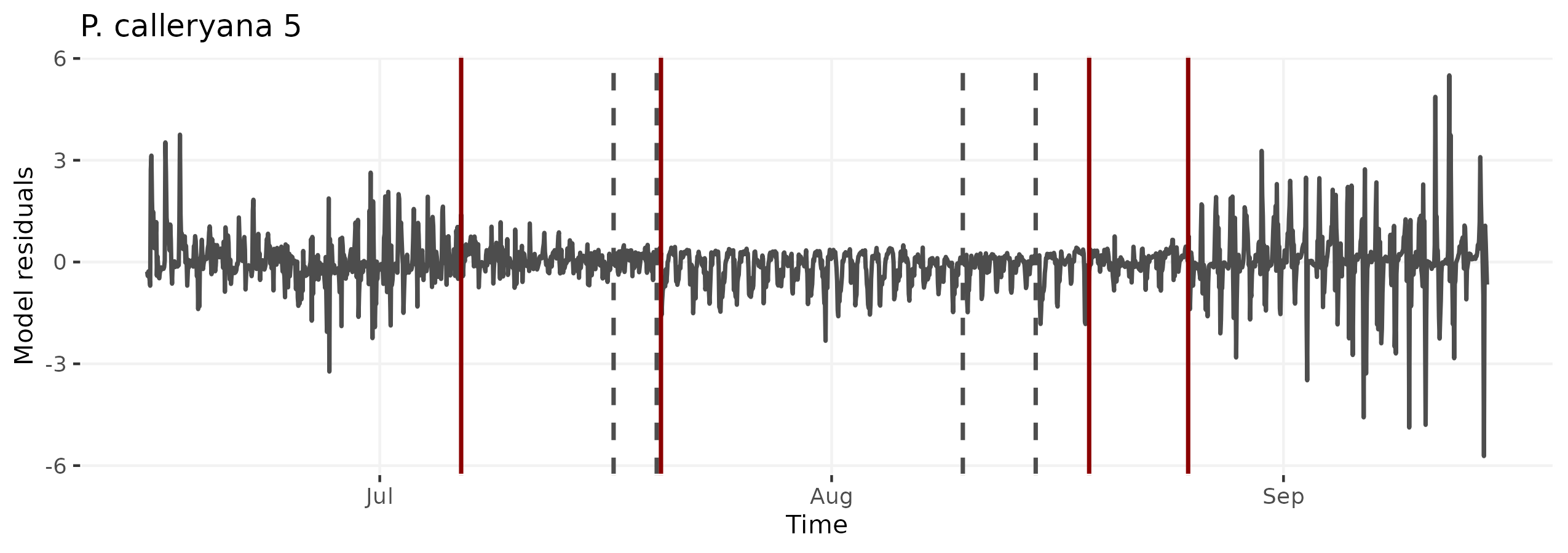}
\captionsetup{labelfont=bf, font=small}
\caption{Segmenting the residual time series of the additive model fitted to \textit{P. calleryana} 1 (top) and \textit{P. calleryana} 5 (bottom) into five segments respectively. The red vertical lines indicate the changepoints and the grey dashed vertical lines indicate the two heatwaves in 2022.}
\label{fig:ResidualCpt}
\end{center}
\end{figure}

The scatter plots in section 4.1 of the main manuscript suggest a potential shift in the water extraction behaviour as a result of the extreme weather condition, e.g., heatwaves. This suggests that using a single model to describe both the normal days and extreme weather days is no longer appropriate. To further investigate this phenomenon, we applied the changepoint detection algorithms on the residual time series from the additive models fitted to the hourly sap flux density of \textit{P. calleryana} 1 and \textit{P. calleryana} 5. Under the null hypothesis that the model is appropriate for the entire time series, the residuals are expected to follow the same distribution. Any significant change in the mean and/or variance of the residual time series can be an indicator that the global model is not suitable for some sections of the time series. Here, the CROPS algorithm \citep{PenaltyCpt} was used to detected multiple changepoints in the time series. The algorithm was implemented using functions in the \texttt{changepoint} package in R \citep{changepoint}. The results from segmenting the residual time series of \textit{P. calleryana} 1 and \textit{P. calleryana} 5 into five segments with four changepoints (indicated by the red solid vertical lines) are presented in Figure \ref{fig:ResidualCpt}. In this case, the number of changepoints are chosen to match the changes in the weather patterns (i.e., pre-heatwaves, heatwave 1, between heatwaves, heatwave 2, post-heatwaves). The results revealed some connections between the changepoint locations and the heatwave periods, which are marked by the grey dashed vertical lines.  

\section{Investigating the impact of size classes}

For commercial tree growers, trees of similar sizes (which is often related to the age of the trees) are often planted and managed together. Hence a practical problem is to estimate the water-use of trees of a particular size. This introduces an interesting question, i.e., whether the size of the trees makes a difference to the relationship between sap flux density and the weather variables and whether this will affect the prediction of sap flux density. In terms of prediction, this translates to whether a model built using data from younger trees can be used to predict the sap flux density of older trees, and vice versa. Here we investigated the impact of size class using the sap flux density time series recorded for the \textit{Carpinus betulus} and the \textit{Ulmus} `New Horizon' of different size classes during the 2022 and 2023 growing seasons. 

\subsection{The impact of size class on the modelled effect} 

Recall that the generic additive model for sap flux density is
\begin{equation}
Y_{t} = \alpha_{0} + \alpha_{1} Y_{t-1} + \alpha_{2} T_{t} + \alpha_{3} H_{t} + \bm{s}_{1}(R_{t})^{\top} \bm{\beta}_{1} + ( \bm{s}_{2}(V_{t}) \cdot R_{t} )^{\top} \bm{\beta}_{2} + \bm{s}_{3}(V_{t} , X_{t})^{\top} \bm{\beta}_{3} + \epsilon_{t} \; .
\label{eqn:GAM_individual_sup}
\end{equation}
To test the impact of size class on the relationship between sap flux density and the weather variables, additive models where the linear term or the smooth non-linear terms interact with size class (as categorical variable) were investigated. Specifically, this was done by introducing the interaction terms, such as $(\bm{s}_{1}(R_{t})^{\top} \otimes C_{i}^{\top}) \; \bm{\beta}_{1}$, $(\bm{s}_{2}(V_{t}, R_{t})^{\top} \otimes C_{i}^{\top}) \; \bm{\beta}_{2}$, where $C_{i}$ is vector of indicator of the size class of tree $i$, and $\bm{\beta}_{1}$, $\bm{\beta}_{2}$ are the augmented coefficient vectors, to the additive models. Transforming size class into a categorical variable enables us to not only test the significance of the variable, but also visualise the impact of size class easily through plotting the interaction terms for each category. We can investigate the differences between the smooth component from different size categories by checking whether the confidence bands or surfaces overlap. 

\begin{figure}[!htb]
\begin{center}
\includegraphics[width=5.2in]{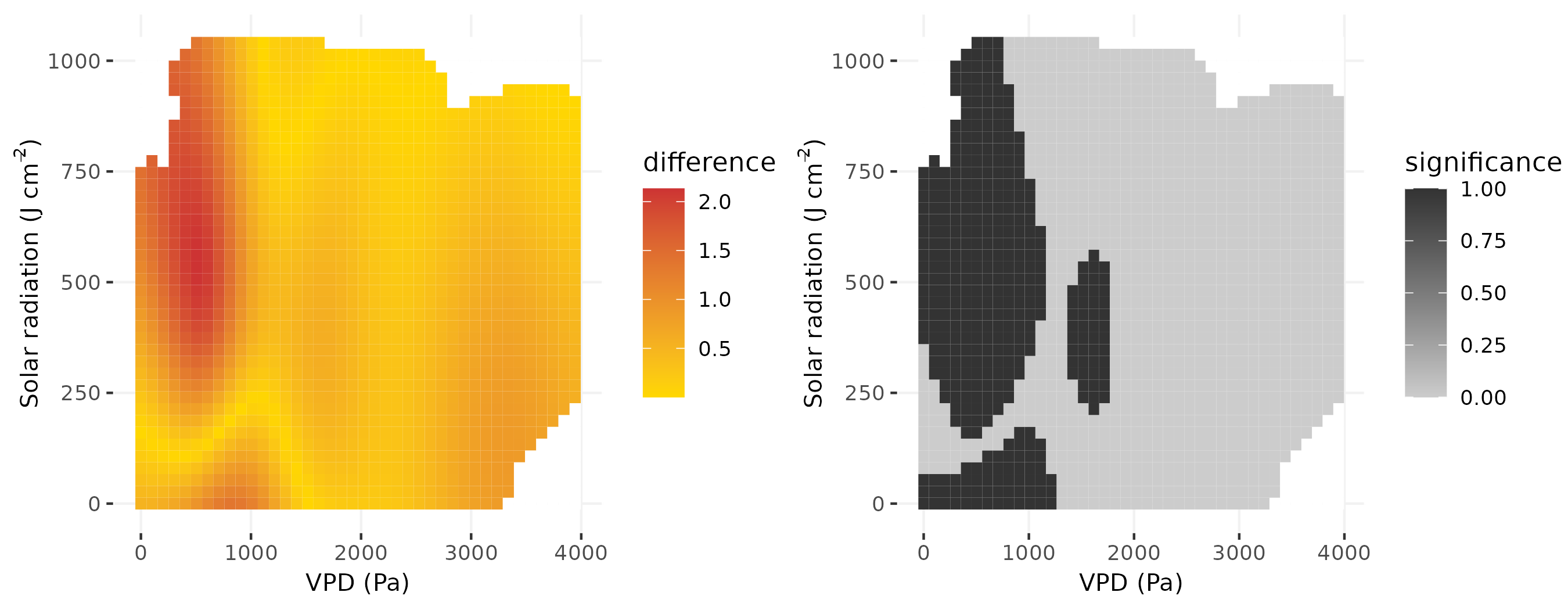}
\includegraphics[width=4.2in]{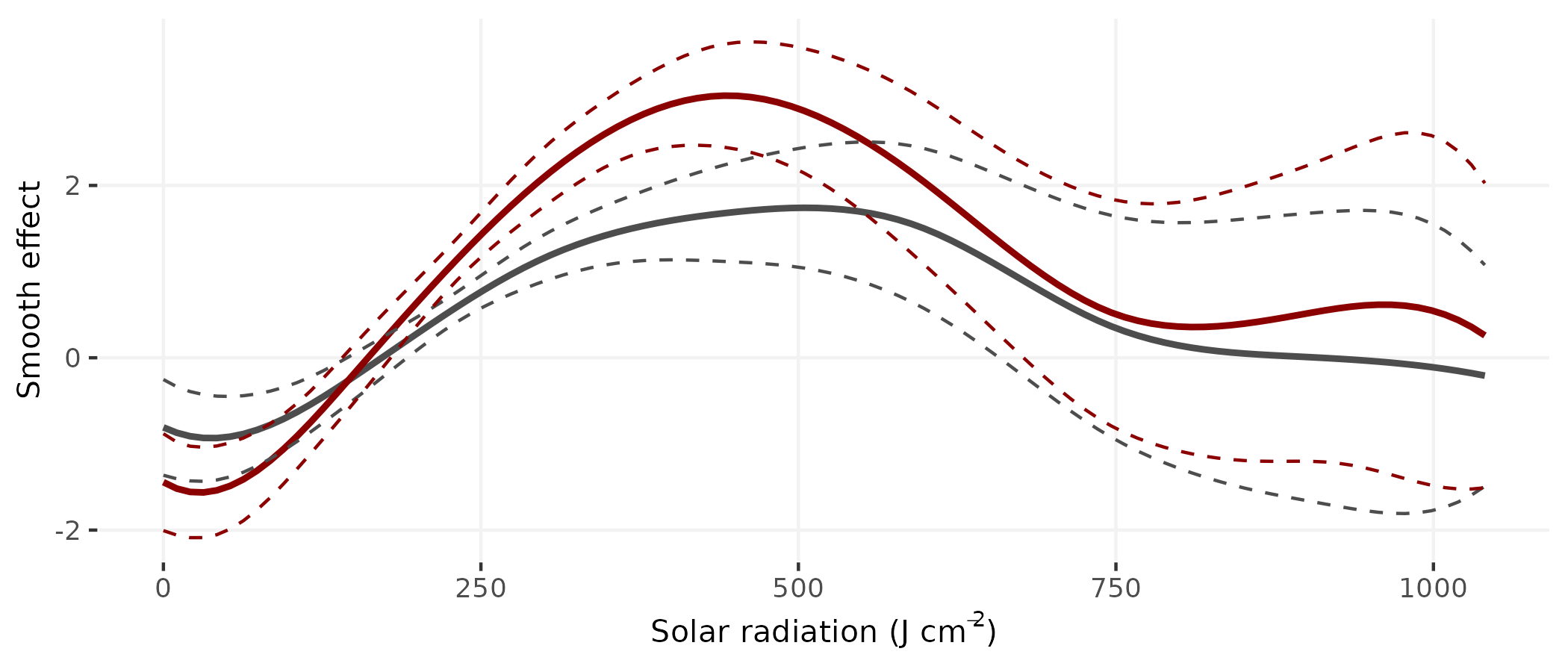}
\captionsetup{labelfont=bf, font=small}
\caption{(Top) Size class difference of type (a) between the \textit{C. betulus} from two size classes. The the heat map show the absolute differences between the estimated smooth surface (large trees - small trees). The map in grey and black show the areas where there are significant differences (at 0.05 level) between the two smooth surfaces. (Bottom) Size class difference of type (b) between the \textit{C. betulus} from two size classes. The solid curves are the smoothed effect and the dashed curves are the 95\% confidence intervals. }
\label{fig:gam_size_carpinus}
\end{center}
\end{figure}

We investigated the impact of size class on 8 large and 5 small \textit{C. betulus} trees and 3 large and 5 small \textit{U.} `New Horizon' trees. Sap flux density time series from the entire growing season after the leave-expansion period were used in this analysis. We tested the following interactions between size class and (a) the interaction surface of VPD and solar radiation, $(\bm{s}_{2}(V_{t}, R_{t})^{\top} \otimes C_{i}^{\top}) \; \bm{\beta}_{2}$, (b) solar radiation $(\bm{s}_{1}(R_{t})^{\top} \otimes C_{i}^{\top}) \; \bm{\beta}_{1}$, (c) the scales of daily cycles, $V_{t} X_{t} C_{i}^{\top} \beta_{3}$, (d) lag-one sap flux density, $Y_{(t-1)i} C_{i}^{\top} \alpha_{1} $, and (e) the combination of (a) and (b). The tests showed significant differences (at 0.05 level) for most of these terms for the trees in \textit{C. betulus} and \textit{U.} `New Horizon' groups. An example showing the differences between the smooth components of two size classes of \textit{C. betulus} are shown in Figures \ref{fig:gam_size_carpinus} 

\subsection{The impact of size class on prediction} 

For comparing the impact of size class on the prediction of sap flux density, we used the superior predictive ability (SPA) test developed by \cite{SPAtest}. The SPA test compares the prediction performance of the benchmark model to that of the competing models to test whether any of the competing models perform better than the benchmark model. Denote the relative performance between the $k$-th model and the benchmark model (referred to as model 0) in terms of predicting time $t$ as
\begin{equation*}
d_{kt} = L_{0}(Y_{t}, \; \hat{Y}_{0t}) - L_{k}(Y_{t}, \; \hat{Y}_{kt}) \; ,
\end{equation*}
for $k = 1, \cdots, m$, where $L_{k}(Y_{t}, \; \hat{Y}_{kt})$ is some loss function, and denote $\bm{d}_{t} = ( d_{1t}, \cdots, d_{mt} )'$, with mean $\bm{\mu} \equiv \mathrm{E}(\bm{d}_{t}) $. The null hypothesis for the test is $H_{0}: \bm{\mu} < 0$. That is, the loss of the competing models are all greater than that of the benchmark model. Under $H_{0}$, the asymptotic distribution of the averaged relative performance, $\bar{\bm{d}} = \frac{1}{n} \sum_{t=1}^{n} \bm{d}_{t}$, is 
\begin{equation*}
n^{1/2} ( \bar{\bm{d}} - \bm{\mu}) \rightarrow \mathbb{N}_{m}(\bm{0}, \bm{\Omega}) \; .
\end{equation*} 
The test statistic used in the SPA test is
\begin{equation}
T_{n}^{SPA} = \max \left\{ \max_{k} \frac{n^{1/2} \bar{d}_{k}}{\hat{\omega}_{k}}, \; 0 \right\} \; ,
\end{equation}
where $\hat{\omega}_{k}$ is the estimated variance of the relative performance of model $k$ to the benchmark model. The critical values of the test are obtained by bootstrapping, where the time series are simulated by randomly drawing segments with lengths following a geometric distribution from the original time series (block bootstrap with random block lengths). If $T_{n}^{SPA}$ goes beyond the critical value, then there is evidence that the competing models perform better than the benchmark. Full detail of the test statistics and the asymptotic results can be found in \cite{SPAtest}. 

To implement the SPA test to our problem, some modifications to the standard testing procedure are required. 

\begin{enumerate}
\item[(a)] To test the difference in predicting a period of time in the future (for example, for the dashboard, we may wish to predict the next 7 days), the stationary bootstrapping procedure described in \cite{SPAtest} and \cite{StationaryBootstrap} needs to be modified to make sure that the re-sampled segments, when connected, are `continuous' in a sense that the bootstrapped time series show similar daily cycles as in the original time series. Therefore, instead of samplling blocks of random lengths following a geometric distribution with a mean of $p$ time points, we sample blocks with lengths being multiples of a day. This can be done by first sampling the lengths of continuous days with a mean of $p^{\ast}$ days, and then connecting the time series of these days at mid-night (00:00 hour each day) when the sap flux density is close to 0. In this way, we ensure the bootstrapped time series are `continuous'. 

\item[(b)] In \cite{SPAtest}, the variance $\hat{\omega}_{k}$ is estimated using the auto-covariance series $\hat{\gamma}_{k}(\tau)$, where $\tau$ is the time lag, from the testing data. Since our testing data come from a few different trees, the estimator of the auto-covariance $\hat{\gamma}_{k}(\tau)$, and hence the estimator of the variance parameter $\hat{\omega}_{k}$, need to be modified. For the testing time series of each tree $i$, we have 
\begin{equation*}
\hat{\gamma}_{k}^{(i)}(\tau) = \frac{1}{n} \sum_{t=1}^{n-\tau} \left(d_{kt}^{(i)} - \bar{d}_{k}^{(i)} \right) \left( d_{k(t+\tau)}^{(i)} - \bar{d}_{k}^{(i)} \right) \, .
\end{equation*}
To obtain $\hat{\gamma}_{k}(\tau)$, we average across $\hat{\gamma}_{k}^{(i)}(\tau)$. The rest remains the same, with
\begin{equation*}
\hat{\omega}_{k}^{2} \equiv \hat{\gamma}_{k}(0) + 2 \sum_{\tau=1}^{n-1} \kappa(n, \tau) \hat{\gamma}_{k}(\tau) \, ,
\end{equation*}
where 
\begin{equation*}
\kappa(n, \tau) \equiv \frac{n-\tau}{n}(1-q)^{\tau} + \frac{\tau}{n}(1-q)^{n-\tau} \, .
\end{equation*}
In the above, $q$ is the parameter of the geometric distribution used in the stationary bootstrap, which may be determined as $q = 1/p$ for a mean segment length $p$. In our application, this would be $q = 1/(p^{\ast} \times 24)$, as it is the number of days that is being sampled. Some discussion on the impact of $q$ is given in \cite{StationaryBootstrap}. 
\end{enumerate}

We ran the SPA tests on the prediction models for \textit{C. betulus} and \textit{U.} `New Horizon' respectively, using the size class information from meta data. The models were built using data up to August of the corresponding growing seasons and they were then used to predict the sap flux density in August. Due to the computational time involved in the bootstrap procedure, we compared the prediction performance of the models with or without size class information within two different prediction windows, each of length 1-week, in August. In this case, the SPA tests showed significant difference in the models for \textit{C. betulus} in the first prediction window (first week of August 2022), but not in the second prediction window (second week of August 2022). For \textit{U.} `New Horizon', including size class information made a significant difference to the predictions of the first week of August 2023, but not the second week of August 2023. These results suggest that the size class can make a significant difference to the prediction of sap flux density. Therefore, it makes sense to consider separating trees from different size class when building prediction models. 

This analysis was implemented in R, with the code modified from the SPA test code from GitHub \url{https://github.com/PedroBSB/mlRFinance/blob/master/R/SPAtests.R}.